\documentclass{iopart}
\usepackage{amsfonts}
\usepackage{amssymb}
\usepackage{amsbsy}
\usepackage{color}
\input epsf.sty

\begin{document}

\def\be{\begin{equation}}
\def\ee{\end{equation}}
\def\ba{\begin{eqnarray}}
\def\ea{\end{eqnarray}}
\newcommand{\BEQ}{\begin{equation}}     
\newcommand{\BEA}{\begin{eqnarray}}
\newcommand{\BD}{\begin{displaymath}}
\newcommand{\EEQ}{\end{equation}}       
\newcommand{\EEA}{\end{eqnarray}}
\newcommand{\ED}{\end{displaymath}}
\newcommand{\eps}{\varepsilon}          
\def\Pr{\mbox{\bf P}}                   
\newcommand{\vph}{\varphi}              
\newcommand{\vth}{\vartheta}            
\newcommand{\D}{{\rm d}}
\newcommand{\II}{{\rm i}}               
\newcommand{\arcosh}{{\rm arcosh\,}}    
\newcommand{\erf}{{\rm erf\,}}          
\newcommand{\wit}[1]{\widetilde{#1}}    
\newcommand{\wht}[1]{\widehat{#1}}      
\newcommand{\lap}[1]{\overline{#1}}     
\newcommand{\demi}{\frac{1}{2}}         
\newcommand{\rar}{\rightarrow}          
\newcommand{\gop}{\wht{\phi}_{\vec{0}}} 

\renewcommand{\vec}[1]{{\bf{#1}}}       
\newcommand{\vekz}[2]
     {\mbox{${\begin{array}{c} #1  \\ #2 \end{array}}$}}
\newcommand{\matz}[4] 
     {\mbox{${\begin{array}{cc} #1 & #2 \\ #3 & #4 \end{array}}$}}
\newcommand{\appsection}[2]{\setcounter{equation}{0}\setcounter{subsection}{0}
\section*{Appendix #1. #2}
\renewcommand{\theequation}{#1.\arabic{equation}}
              \renewcommand{\thesection}{#1} }
\newcommand{\appsektion}[1]{\setcounter{equation}{0}\setcounter{subsection}{0}
\section*{Appendix. #1}
\renewcommand{\theequation}{A.\arabic{equation}}
              \renewcommand{\thesection}{A} }
\title{Exact correlation functions in particle-reaction models with immobile particles}

\author{Christophe Chatelain,$^a$ Malte Henkel,$^a$ 
M\'ario J. de Oliveira$^b$ and T\^ania Tom\'e$^b$}

\address{$^a$Groupe de Physique Statistique, \\
D\'epartement de Physique de la Mati\`ere et des Mat\'eriaux, \\
Institut Jean Lamour (CNRS UMR 7198), 
Universit\'e de Lorraine Nancy, \\
B.P. 70239, F - 54506 Vand{\oe}uvre les Nancy Cedex, France\\~\\
$^b$ Instituto de F\'{\i}sica, Universidade de S\~ao Paulo, C.P. 66318, \\
05314-970 S\~ao Paulo SP, Brasil
}
\ead{}

\begin{abstract}
Exact results on particle-densities as well as correlators in two models of immobile particles, 
containing either a single species or else two distinct species, are derived.
The models evolve following a descent dynamics through pair-annihilation where each 
particle interacts at most once throughout its entire history. The resulting large number
of stationary states leads to a non-vanishing configurational entropy. 
Our results are established for arbitrary initial conditions and are derived 
via a generating-function method. 
The single-species model is the dual of the $1D$ zero-temperature kinetic Ising model with
Kimball-Deker-Haake dynamics. In this way, both infinite and semi-infinite
chains and also the Bethe lattice can be analysed. 
The relationship with the random sequential adsorption of
dimers and weakly tapped granular materials is discussed. 
\end{abstract}


\section{Introduction} 

Exactly solvable models play an important role in understanding the complex behaviour
of strongly interacting many-body systems, 
see \cite{Priv97,Marr99,benA00,Schu00,Tome01,Henk08} and refs. therein.
With very few exceptions, exactly solvable models are largely found in one spatial dimension. 
Exact solutions are important because standard mean-field schemes are inadequate in low
dimensions where effect of fluctuations is strong, as confirmed experimentally.
Classic examples are the kinetics of excitons on long chains of the polymer chain
TMMC = (CH$_3$)$_4$N(MnCl$_3$) \cite{Kroo93}, 
and other polymers confined to quasi-one-dimensional geometries
\cite{Pras89,Kope90}. Systems like carbon nanotubes, which are essentially one-dimensional,
have been extensively studied; for instance, the relaxation of photo-excitations \cite{Russ06} 
or the photoluminescence saturation \cite{Sriv09}.
What is common in all these systems is that particles move diffusively 
on an effectively one-dimensional lattice, and when two particles encounter, one of them
disappears with probability one.  These models can be solved,
for example by the well-established method of empty intervals \cite{benA00,benA07,Agha05}. 
The average particle concentration $c(t) \sim~ t^{-1/2}$ \cite{Tous83}
found by this method is in complete agreement with the experimental results,
unlike the mean-field prediction $c_{\mbox{\rm\footnotesize MF}}(t) \sim t^{-1}$. 

\begin{figure}[tb]
\centering
\centerline{\epsfxsize=6.5cm\epsfbox{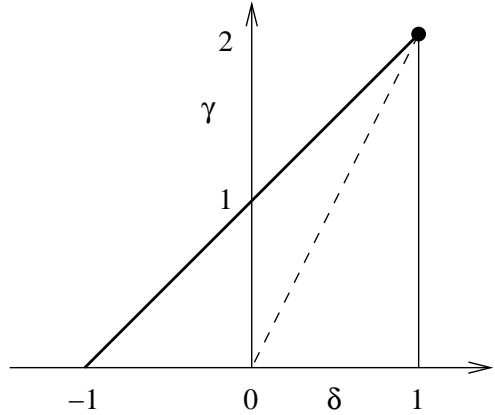} }
\caption[fig0]{Parameter space of kinetic Ising chains as parametrised by $\gamma$ and $\delta$ in 
(\ref{2}). The usual Glauber dynamics corresponds to $\delta=0$
and the {\sc kdh} dynamics is given by $\gamma=2\delta$ (dashed line).
The zero-temperature line corresponds to $\gamma=1+\delta$
(thick line) and the infinite-temperature line to $\gamma=0$. 
The full circle at $\delta=1$ and $\gamma=2$ corresponds to
the dynamics analysed here. \label{kdh_fig0}}
\end{figure}

Besides particle systems, one may consider the kinetics of magnetic systems
whose phenomenology can be captured by kinetic Ising models \cite{Glauber63}. 
We point out that such models are relevant for the description of 
the slow relaxation dynamics of real systems 
such as the single-chain magnet Mn$_2$(saltmen)$_2$Ni(pao)$_2$(py)$_2$](ClO$_4$)$_2$ \cite{Coul04}. 
The configurations of the Ising chain are $\{\sigma\} = (\sigma_1,\sigma_2,\ldots\sigma_L)$, 
with the Ising spins $\sigma_{i}=\pm 1$ attached to each site $i$ of the chain. 
The master equation describes
the sequential time-evolution of the probability $\Pr(\{\sigma\};t)$ of a configuration $\{\sigma\}$
\BEQ \label{1}
\hspace{-1.8truecm}{\partial\over \partial t}\Pr(\{\sigma\};t)
   =\sum_i\big[\omega_i(-\sigma_i)\Pr(\ldots -\sigma_i\ldots;t)
   -\omega_i(\sigma_i)\Pr(\ldots \sigma_i\ldots;t)\big]
\EEQ 
where the \ldots means that all other spins are not affected and the
choice of the rates $\omega_i(\sigma_i)$ selects a specific model.
The most general form of the transition rates for the Ising chain 
which does not impose explicit conservation laws, keeps the global $\mathbb{Z}_2$-symmetry $\sigma_i \mapsto - \sigma_i$ 
and which only takes into account the spin 
$\sigma_i$ and its two nearest neighbours $\sigma_{i\pm 1}$, is given by
\begin{equation} \label{2}
\omega_i(\sigma_i) = \frac12\left[1-\frac{\gamma}2\sigma_i
(\sigma_{i-1}+\sigma_{i+1}) + \delta \sigma_{i-1}\sigma_{i+1}  \right]
\end{equation}
where the parameters $\gamma$ and $\delta$ are restricted to
the intervals $0\leq\gamma\leq 1+\delta$ and $|\delta|\leq 1$. This parameter space
is illustrated in figure~\ref{kdh_fig0}. 
One requires these transition rates to obey the {\em detailed balance condition} 
\begin{equation} \label{bd}
\omega_i(\sigma_i)\Pr_{\rm eq}(\ldots\sigma_i\ldots) 
= \omega_i(-\sigma_i)\Pr_{\rm eq}(\ldots-\sigma_i\ldots)
\end{equation}
where $\Pr_{\rm eq}(\ldots\sigma_i\ldots) = Z^{-1} \exp[-{\cal H}[\sigma]/T]$ 
is the equilibrium probability distribution where $T$ denotes the temperature of the heat bath and  
${\cal H}=-J\sum_i\sigma_i\sigma_{i+1}$ is the Ising chain Hamiltonian
with an exchange coupling $J$. Then detailed balance implies
\begin{equation}
\gamma = (1+\delta)\tanh(2J/T)
\end{equation}
The Glauber-Ising model is specified by the choice \cite{Glauber63} 
\begin{equation}
\delta=0 \;\; , \;\; \gamma=\tanh(2J/T)
\end{equation}
Here, we shall rather consider an alternative form, proposed by Kimball \cite{Kimball79} and Deker and 
Haake \cite{Deker79} ({\sc kdh}), namely
\begin{equation} \label{kdh}
\delta = \frac{\gamma}{2} =  \frac{\tanh(2J/T)}{2-\tanh(2J/T)}
\end{equation}

We remark that along the zero-temperature line $\gamma=1+\delta$ 
the detailed balance condition (\ref{bd})
is trivially obeyed, since both sides of equation (\ref{bd})
vanish (for $T=0$ all stationary states are absorbing states).

Physically, the Glauber-Ising and {\sc kdh} models are very different. 
This may be understood first from the exact relationship between the relaxation time $\tau$ and
the spatial equilibrium correlation length $\xi$, viz. 
\BEQ
\tau \sim \xi^z \;\; , \;\; z = \left\{
\begin{array}{llr} 2 & \mbox{\rm ~~;~ Glauber}   & \cite{Glauber63} \\
                   4 & \mbox{\rm ~~;~ {\sc kdh}} & \cite{Deker79,Haake80,Dutta09}
\end{array} \right.
\EEQ
which in both models can be established from the exactly calculable time-dependent average spin. 
Second, and more crucially, at $T=0$, the
Glauber-Ising model has only two absorbing states (a feature typical of a relaxing simple magnet), 
whereas the {\sc kdh}-model has $\sim\left( (1+\sqrt{5\,})/2\right)^L$ \cite{Carl01,deSm02} absorbing
states on a chain of $L$ sites (a feature more typical of a relaxing glassy system, 
such as the Frederikson-Andersen model \cite{Fred84}). While almost any quantity of
interest can be computed explicitly in the $1D$ Glauber-Ising model, see e.g. \cite{Glauber63,Priv97}, 
for the {\sc kdh}-model only global averages \cite{Kimball79,Deker79}
or certain global correlators and responses have been found \cite{Dutta09}. 
In the {\sc kdh}-model, often the corresponding equations of motion
only close for certain classes of initial conditions and in the $L\to\infty$ limit \cite{Dutta09}. 

Our work started from an attempt to find more exactly computable averages 
for the {\sc kdh}-model. For the remainder of this paper,
we shall restrict to zero temperature $T=0$, hence $\delta=1$ and $\gamma=2$ for the {\sc kdh}-model. 
Then one can define the dual kink variable $\eta_i\in\{0,1\}$ and the rates (\ref{kdh}) become
\BEQ \label{4}
\hspace{-1.8truecm}\eta_i:={1-\sigma_i\sigma_{i+1}\over 2} \;\; , \;\; 
\omega_i(\sigma_i)=     
     {1\over 2}\big(1-\sigma_i\sigma_{i-1}\big)
     \big(1-\sigma_i\sigma_{i+1}\big)   
     =2\eta_{i-1}\eta_i
\EEQ 
Single-spin flips $\sigma_i\rightarrow -\sigma_i$ are associated
to the annihilation or creation of a pair of kinks
\BEQ
     \eta_{i-1}\longrightarrow 1-\eta_{i-1} \;\; , \;\;
     \eta_i\longrightarrow 1-\eta_i
\EEQ
The only allowed transition is thus the annihilation of a pair
of kinks. Neither creation nor diffusion of kinks are allowed. 
Therefore, {\em the $1D$ {\sc kdh}-model at $T=0$
is dual to a model where the sites are either empty or else 
occupied by a single immobile particle of a single species $A$, 
which can undergo pair-annihilation 
$A+A\to\emptyset+\emptyset$ with their nearest neighbours and with rate $2$}.

Besides being of interest in its own right as an abstract problem in many-body physics, there exist several
physical motivations for the study of this model:
\begin{enumerate} 
\item \underline{\em Random sequential adsorption} ({\sc rsa}): 
In the usual {\sc rsa} defined on a regular lattice, an atom
adsorbed in a site excludes the adsorption of atoms in
the neighbouring sites. Instead of single atoms one may define
{\sc rsa} of dimers on a regular lattice in which the two atoms of
a dimer occupy two nearest-neighbour sites of the lattice. 
In one dimension the two versions are equivalent. One then investigates quantities such as the density of
particles in the stationary state \cite{Evan93}.
If we define the variable $\overline{\eta}_i$
that takes the value 0 or 1 according to whether the site $i$
is empty or occupied by on atom of the dimer, then the relation between
this variable and the kink variable $\eta_i$ is simply
$\overline{\eta}_i=1-\eta_i$ and the transition rate of 
the {\sc rsa}-model is that given by equation (\ref{4}). A kink ($\eta_i=1$)
corresponds therefore to an empty site ($\overline{\eta}_i=0$)
in the {\sc rsa}-formulation. Notice that not any state
$\{\overline{\eta}_i\}$ is a possible configuration of dimers. However,
if the initial state is a dimer configuration then any future
state will also be a dimer configuration.
Random sequential adsorption is a commonly encountered mechanism when
macromolecules or collo\"{\i}dal particles deposit on solid surfaces 
and when gravity effects can be neglected. 
It goes much beyond the often inadequate mean-field-like descriptions based on the Langmuir equation. 
See \cite{Scha00,Rabe11} for recent reviews, including experimental tests and applications.  
\item \underline{\em Granular matter}: 
the particles making up granular materials rapidly relax into one of the
very many blocked configurations of these systems. Under the effect of a gentle {\em tapping}, 
they may jump from one
of these blocked states to another. 
These question has been raised how to formulate a statistical description of granular
matter in terms of an ensemble of these blocked states. 
The {\em Edwards hypothesis} states that all blocked  configurations
(`valleys') with the same energy should be equivalent 
and that their number can be counted simply in terms of the
configurational entropy (or  `complexity') \cite{Edwa94}. 
While this proposal seems to work very well for mean-field-like models,
one of the interest of exactly solvable models such as the ones 
considered here is that because of their large number of
stationary states, they have a non-vanishing complexity and hence permit a precise test of these
concepts without appealing to mean-field schemes. 
Indeed, it has been shown that although the Edwards proposal is numerically quite accurate, it 
leads to predictions different from what is found in exactly solved one-dimensional systems 
\cite{Prad01,Prad02,deSm02,Brey03,Tarj04,Brey07}. 
For reviews, see e.g. \cite{Godr05,Dauch07}. 
\end{enumerate}
Therefore, we shall present here exact derivations for particle-densities and some correlators, 
for two models undergoing a 
`descent dynamics' where each individual particle only reacts {\em once} through the entire history. 
The first one simply is
the pair-annihilation model $A+A\to\emptyset+\emptyset$ 
of diffusion-less particles, with the rate (\ref{4}), and dual to the Ising model with {\sc kdh}-dynamics, 
see sections~2 and~3.
The second model describes the pair annihilation 
$A+B\to \emptyset+\emptyset$ of two distinct species of diffusion-less particles, see
section~4. A possible application of the two-species model is related to the fact that it could be used
to distinguish the two types of kinks in the Ising model, namely 
$\uparrow\downarrow\, \doteq A$ and $\downarrow\uparrow\, \doteq B$, such that one would have the
duality correspondences 
\BD
\left( \uparrow\downarrow\uparrow \longrightarrow \uparrow\uparrow\uparrow\right) \doteq 
\left( AB \longrightarrow \emptyset\emptyset\right)
\;\; , \;\;
\left( \downarrow\uparrow\downarrow \longrightarrow \downarrow\downarrow\downarrow\right) \doteq 
\left( BA \longrightarrow \emptyset\emptyset\right)
\ED

While the reaction-diffusion processes mentioned above can be treated via the 
{\em empty-interval method}, the diffusion-free models
we shall be considering here can be treated by its converse \cite{Evan93}, 
which is sometimes called the {\em full-interval method} \cite{Khor12}. 
A central quantity will be 
$F_n := \Pr\big(\underbrace{\bullet \bullet \ldots \bullet}_{n \mbox{\rm\footnotesize ~particles}}\big)$, 
the probability of having at least $n$ consecutive sites occupied by a particle. 
It turns out that the $F_n$ satisfy closed linear systems of equations
of motion, to be derived from (\ref{4}).\footnote{In contrast with the 
full-interval method as used in \cite{Khor12,Agha05}, we do {\em not}
require the condition $F_0(t)=1$, but shall rather analytically continue 
the equations of motion to the case $n=0$ which will then be used
to {\em define} $F_0(t)$ by this differential equation. 
In this respect the full-interval and empty-interval methods are quite distinct,
although the respective equations of motion look very similar.} 
Here, we shall go beyond the calculation of mere densities. Since in the
context of the empty-interval method, an efficient way to 
generate correlators and responses is to consider probabilities of pairs
of empty sites \cite{Doer90,benA98,benA00,deSm02,Dura10,Dura11}, 
we shall derive correlators by solving the equations of motion of the probabilities of
pairs of groups of occupied sites, separated by a gap. 
Rather than restricting to rather special initial conditions such that
the solutions of motion can be guessed by making an appropriate ansatz, 
we shall present a very general and easy-to-use generating-function method which allows to derive systematically all 
quantities of interest for {\em arbitrary} initial conditions. The usually considered situation
of initially uncorrelated particles is recovered as a special case. 

Binary {\sc rsa} reactions of immobile point-like particles have also been studied since a long time in the {\em continuum}. Besides
the so-called `black spheres model', where particles closer than a given distance $r_0$ react instantaneously \cite{Kuzo88},
other commonly studied interactions include `recombination through tunnelling', where the reaction probability
$w(r) = w_0 \exp(- r/r_0)$ depends exponentially on the distance $r$ between the interacting particles or
`multi-polar interactions' with the form $w(r) = w_0 (r/r_0)^{-s}$. In these cases, reactions become possible across
arbitrarily large distances which leads to a totally different structure of stationary states than the one of the
lattice models treated in our work. In view of the effective long-range interactions thereby induced between all particles, 
it is not totally surprising that mean-field-like approximations, such as Kirkwood's `superposition approximation' \cite{Kirk35,Bale75}
(which serves to close the infinite hierarchy of equations of motion for densities, pair correlators,\ldots 
and is quite similar in spirit to the 
`pair-approximation' or `(2,1)-approximation' in the systematic analysis in \cite{benA92}), 
produce numerically very precise results \cite{Burl87,Schn89,Schn90,Ludi91,Kapr92} and which have been confirmed later
by precise analytical studies \cite{Burl90,Bonn95,Bonn00}. Generically, only the case of initially uncorrelated particles
(and of equal initial density $\rho_A(0)=\rho_B(0)$ in the two-species model) is considered, and one obtains for 
very long times $t\to\infty$ a very slow decrease of the average
particle density $\rho_A(t) \sim \ln^{-d} t$ for the single-species model and $\rho_A(t) = \rho_B(t) \sim \ln^{-d/2} t$ for
the two-species model, both with recombination through tunnelling in $d$ spatial dimensions 
\cite{Schn89,Schn90,Bonn95} and furthermore dynamical
scaling forms for the correlators \cite{Bonn00}. For multi-polar interactions, one obtains $\rho_A(t)=\rho_B(t)\sim t^{-d/(2s-d)}$
\cite{Burl90,Ludi91}. However, these models with the Kirkwood approximation do not preserve the exponentially growing number of
steady states with the system size $L$ (of the order $\sim 1.618^L$ for the single-species model and $\sim 2.41^L$ for the two-species model)
of our {\em discrete} lattice models with contact interactions only
and we do not need to fall back on any approximation scheme.

This paper is organised as follows. 
In section~2, we consider the diffusion-free pair annihilation model 
with the simplifying technical
assumption of spatial translation-invariance. 
Particle densities and pair correlators are derived exactly, for arbitrary
initial conditions, via a generating function method. 
We shall study both the cases of an one-dimensional chain \cite{Kemk81,Oliv94,Prad01,deSm02} 
and as well the Bethe lattice \cite{Evan84,Maju93,Abad04} 
as an example of an effectively infinite-dimensional lattice. 
In section~3, we generalise to the case where spatial translation-invariance is no longer required. 
In particular, our treatment includes the case of an one-dimensional, semi-infinite system
so that one can study the cross-over from the boundary towards the bulk. 
In section~4, we illustrate further the versatility of 
our generating function technique by deriving densities \cite{Maju93} 
and correlators of a two-species pair-annihilation model of immobile species
$A,B$ of particles undergoing the reaction $A+B\to\emptyset+\emptyset$. 
Section~5 gives our conclusions. The computational details for the single- and two-species models are presented in 
the appendices~A, B and~C. Appendix~D outlines the enumeration 
of stationary states in the two-species model.

\newpage
\section{The pair-annihilation model without diffusion}

Consider the diffusion-free pair-annihilation model, consisting of particles of a 
single species $A$, which sit motionless on the sites $i$ 
of an infinite linear chain and which may undergo the only admissible
reaction $A+A\to\emptyset+\emptyset$ with rate $2$ between nearest-neighbour sites. 
{}From the master
equation (\ref{1}), with the rates (\ref{4}), 
one can derive the equations of motion for the averages of interest. 
In general, any average $\langle f(\{\eta\})\rangle$ satisfies the evolution equation
\BEA
    \hspace{-2.0truecm}\lefteqn{{\D\over \D t}\langle f(\{\eta\})\rangle 
    =\sum_{\{\eta\}} f(\{\eta\}){\partial\over \partial t}\Pr(\{\eta\};t)}
    \nonumber\\
    \hspace{-2.0truecm} &=& 2\sum_i \sum_{\{\eta\}} f(\{\eta\}) \Big[(1-\eta_{i-1})(1-\eta_i)
    \Pr(\ldots(1-\eta_{i-1}),(1-\eta_i)\ldots;t)\Big. 
    \nonumber\\
    \hspace{-2.0truecm}& & \Big. \hspace{2.0truecm}
    -\eta_{i-1}\eta_i\Pr(\ldots\eta_{i-1},\eta_i\ldots;t)\Big]\nonumber\\
    \hspace{-2.0truecm}&=& 2\sum_i \sum_{\{\eta\}}
    \big[f(\ldots(1-\eta_{i-1}),(1-\eta_i)\ldots)-f(\{\eta\})\big]
    \eta_{i-1}\eta_i\Pr(\ldots\eta_{i-1},\eta_i\ldots;t)
    \nonumber\\
    \hspace{-2.0truecm}&=&2\sum_i \langle \big[f(\ldots(1-\eta_{i-1}(t)),(1-\eta_i(t))\ldots)
    -f(\{\eta\})\big]\eta_{i-1}(t)\eta_i(t)\rangle
    \label{EqEvolution}
\EEA

\subsection{Density of $n$-strings }

We shall concentrate on {\em $n$-strings} of at least $n$ consecutive particles 
$\underbrace{\bullet\bullet\ldots\bullet}_{\mbox{\rm\footnotesize $n$ particles}}$ 
(or kinks in the spin formulation of the {\sc kdh}-model at $T=0$). Their time-dependent density is 
\BEQ 
    C_n(t):=\langle \eta_1(t)\eta_2(t)\ldots\eta_n(t)\rangle
\EEQ
According to (\ref{EqEvolution}), the $C_n(t)$ satisfy the equations of motion 
    \ba{\D \over \D t}C_n(t)
    =&-2\langle \eta_0(t)\eta_1(t)\eta_2(t)\ldots\eta_n(t)\rangle\nonumber\\
    &-2(n-1)\langle \eta_1(t)\eta_2(t)\ldots\eta_n(t)\rangle\nonumber\\
    &-2\langle \eta_1(t)\eta_2(t)\ldots\eta_n(t)\eta_{n+1}(t)\rangle
    \label{eqev1}\ea
where we used the relations $\eta_i^2=\eta_i$ and
$(1-\eta_i)\eta_i=0$ which follow since $\eta_i\in\{0,1\}$.

\subsubsection{General solution}
Throughout this section, we shall assume that the initial conditions display translation-invariance. 
Since we shall later follow essentially the same approach towards finding the solution, we explain it
here, for the most simple case, in a little more detail, 
although the end result for the $n$-string density $C_n(t)$ is well-known, 
at least for uncorrelated initial conditions \cite{Kemk81,Evan84,Evan93,Oliv92,Oliv94,Prad01,Tome01,deSm02}. 
The evolution equation (\ref{eqev1}) now becomes, for $n\geq 1$
   \be {\D \over \D t}C_n(t)=-4C_{n+1}(t)-2(n-1)C_n(t) \label{eqEvolCn}\ee
To eliminate the last term in (\ref{eqEvolCn}), let
    \be C_n(t)=U_n(t)e^{-2(n-1)t}   \label{Trick1}\ee
so that we are left with
    \be {\D \over \D t}U_n(t)=-4U_{n+1}(t)e^{-2t} \label{EqEvolun}\ee
The exponential factor can be removed by setting $U_n(t)=u_n(s)$, where 
    \be s={e^{-2t}-1\over 2}\ \Leftrightarrow
    \ t=-{1\over 2}\ln(2s+1)  \label{RenorTime}\ee
so that, for all $n\geq 1$ 
    \be {\D\over \D s}u_n(s)=4u_{n+1}(s)\label{EqEvolun2}\ee
In order to simplify the following computation, and all those shall follow later, 
we now define an auxiliary quantity $u_0(s)$ such that
(\ref{EqEvolun2}) is valid for all $n\geq 0$. 
We shall show below that $u_0(s)$ 
does not enter explicitly into any physical observable of interest. 
Equation (\ref{EqEvolun2}) 
is now solved by introducing the generating function
   \be F(x,s) :=\sum_{n=0}^{\infty} {u_n(s)\over n!}x^n
   \label{DefF}\ee
which in turn satisfies the equation 
    \be \hspace{-1.0truecm}{\partial\over \partial s}F(x,s)
    =\sum_{n=0}^{\infty} {4u_{n+1}(s)\over n!}x^n
    =4\sum_{n=1}^{\infty} {u_{n}(s)\over (n-1)!}x^{n-1}
    =4{\partial\over \partial x}F(x,s)\ee
which has the general solution $F(x,s)=f(x+4s)$. 
Herein, the last unknown function $f$ 
is fixed from the initial condition $f(x)=F(x,0)$. We finally have
   \be \label{eq20}
   F(x,s)=f(4s+x) = F(x+4s,0) 
   \ee
Since the generating function $F$ is analytic at $x=0$ by construction, 
we can expand it according to (\ref{DefF}) and  find
   \ba 
   F(x,s)&=\sum_{k=0}^{\infty} {u_k(0)\over k!}(4s+x)^k\nonumber\\
   &=\sum_{k=0}^{\infty} {u_k(0)\over k!}\sum_{n=0}^k
   {k!\over n!(k-n)!}(4s)^{k-n}x^n\nonumber\\
   &=\sum_{n=0}^{\infty}\sum_{k=n}^{\infty} u_k(0)
   {(4s)^{k-n}\over n!(k-n)!}x^n\nonumber\\
   &=\sum_{n=0}^{\infty} \frac{1}{n!}\left( \sum_{m=0}^{\infty}{u_{m+n}(0)\over m!}(4s)^m \right) x^n 
   \label{FctGeneF1}
   \ea
where we also exchanged the order of summation and performed a change of variable in the index $k$. 
Comparing coefficients of $x$, it follows that for all $n\geq 0$
   \be u_n(s)=\sum_{m=0}^{\infty} {u_{m+n}(0)\over m!}(4s)^m\label{Soluns}\ee
Reconverting this to the function $U_n(t)=u_n(s)$, the $n$-string density $C_n(t)$ is finally
\BEQ 
   C_n(t)=\sum_{m=0}^{\infty} {2^m\over m!}C_{m+n}(0)
   \left(e^{-2t}-1\right)^m \; e^{-2(n-1)t}  \label{SolFinCorr}
\EEQ
since consideration of the $t\to 0$ limit shows that 
$u_n(0)=C_n(0)$ for all $n\geq 0$.
Clearly, the physically relevant $n$-string densities 
$C_{n}(t)$ with $n\geq 1$ are independent of the
initial value $C_0(0)$, as it should be. 
Eq.~(\ref{SolFinCorr}) will become an important initial condition when we
shall compute correlators below. 

\subsubsection{Specific initial conditions}
Eq.~(\ref{SolFinCorr}) is our first result, and gives 
$C_n(t)$ in terms of the all initial $n$-string densities $C_n(0)$. 
In order to illustrate the physical content, 
we consider the case of initially totally uncorrelated particles, with
average density $\rho$, such that, for $n\geq 0$
\BEQ
C_n(0) = \rho^n
\EEQ
Carrying out the sum in (\ref{SolFinCorr}) is straightforward, hence 
\BEQ \label{21}
    C_n(t)=\rho^n e^{2\rho\big(e^{-2t}-1\big)-2(n-1)t}
\EEQ
and shows the characteristic double exponential in the time-dependence 
\cite{Kemk81,Evan93,Prad01,Tome01,deSm02}.\footnote{We point out an
important difference with the fairly similar empty-interval method, 
which considers the probability $E_n(t)$ of finding $n$ 
consecutive empty sites, see \cite{benA00}. If in addition one assumes that $E_0(t)=1$, 
the $E_n(t)$ obey equations of motion  quite similar to
(\ref{eqEvolCn}) and which are consistent with $E_0(t)=1$ remaining valid for all times \cite{Dura10}. 
However,  for the diffusion-less case under study in this paper, even if one starts from a fully occupied
lattice initially, viz. $\rho=1$ and admits $C_n(0)=1$ $\forall n\geq 0$, 
one has $C_0(t)=e^{2t-2(1-e^{-2t})}\ne 1$.} 

\begin{figure}[tb]
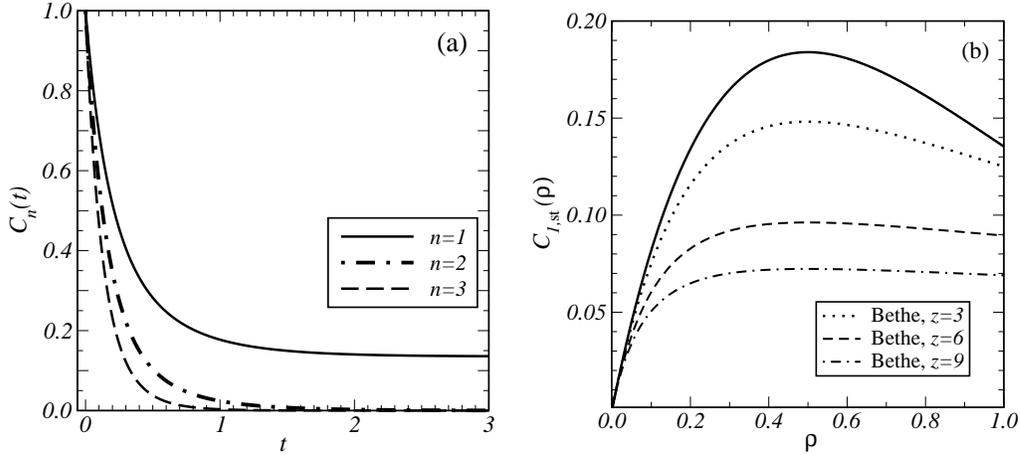

\centerline{\epsfxsize=6.5cm\epsfbox{kdh_fig1a.eps} ~~ \epsfxsize=6.5cm\epsfbox{kdh_fig1b.eps} }
\caption{\label{fig1} (a) Time-dependence of the probability 
$C_n(t)$ of at least $n$ consecutive occupied sites,
for $n=[1,2,3]$ from top to bottom and for an initially fully occupied lattice with $\rho=1$. 
(b) Stationary particle-density $C_{1,\mbox{\rm st}}(\rho)=\lim_{t\to\infty} C_{n}(t)$ 
as a function of
the initial density $\rho$ (full line). The dashed/dotted lines gives the stationary 
particle-density on the Bethe lattice for several coordination numbers $z$. 
}
\end{figure}

Two values of $\rho$ have a particular interpretation for the {\sc kdh}-model at $T=0$. 
\begin{enumerate}
\item[$\rho=\demi$:] this corresponds to random initial conditions for the Ising spins 
so that the probability of a kink is $\demi$.
\item[$\rho=1$:] this corresponds to an ordered anti-ferromagnetic initial state 
$\sigma_i=(-1)^i$, and all bonds are occupied by a kink. 
\end{enumerate}
In figure~\ref{fig1}a, we show the time-evolution of the $n$=string density. 
While for $n=1$, the density rapidly converges towards a finite
value in the limit of large times $t\to\infty$, larger $n$-strings 
(with $n\geq 2$) do not survive. In figure~\ref{fig1}b, we show
the dependence of the stationary density 
$C_{1,\mbox{\rm\footnotesize st.}}(\rho)=C_1(\infty)=\rho e^{-2\rho}$ 
on the density $\rho$. The decrease of
that density when $\rho\to 1$ is intuitively accounted for by 
observing that for $\rho$ large, many more particles will find a partner
for reaction whereas for smaller values of $\rho$, more isolated particles will remain. 
The maximal stationary density is achieved for
$\rho_m=\demi$, where $C_{1,\mbox{\rm\footnotesize st.}}(\demi)=1/(2e)\simeq 0.1839\ldots$.  

Usually the initial condition for the {\sc rsa} of dimers
is taken as an empty lattice which corresponds to
all bonds occupied by a kink. In this case $C_n(0)=1$ 
and the well-known results for {\sc rsa} in one
dimensions can be off from (\ref{21}) with $\rho=1$, 
including the stationary density $C_1(\infty)=e^{-2}$. 
It is worth to consider distinct initial conditions for the {\sc rsa} of dimers. 
For instance, consider a configuration of dimers constructed by placing dimers at a chain 
as follows. Let us denote by AB a dimer so that a  
configuration of dimer will be 
$ABAB\emptyset AB\emptyset\emptyset AB\ldots$ where $\emptyset$ denotes an empty site.
Walking along the sites of the chain, starting from
the origin, if site $i-1$ is empty or occupied by a B atom
the probability of site being occupied by an A atom is $p$
and to remain empty is $q=1-p$. If site $i-1$ is A, then site
$i$ will be B. This defines a Markov chain whose stationary
solution is such that the density of vacant sites is $\rho=q/(2-q)$
and the probability of a sequence of $n$ empty sites
is $C_n(0)=q^n/(2-q)$ which 
we consider to be the initial condition. From equation (\ref{SolFinCorr}), it follows
\begin{equation}
\hspace{-0.5truecm}C_n(t) = \frac{q^n}{2-q} \exp\left[ {2q(e^{-2t}-1)-2(n-1)t}\right] 
\mbox{\rm ~~~where~~} q=\frac{2\rho}{1+\rho}
\end{equation}
so that $C_1(\infty) = qe^{-2q}/(2-q)=\rho e^{-4\rho/(1+\rho)}$
which is a monotonically increasing function of $\rho$.
The usual initial condition for dimers is recovered for $q=1$.

\begin{figure}[tb]
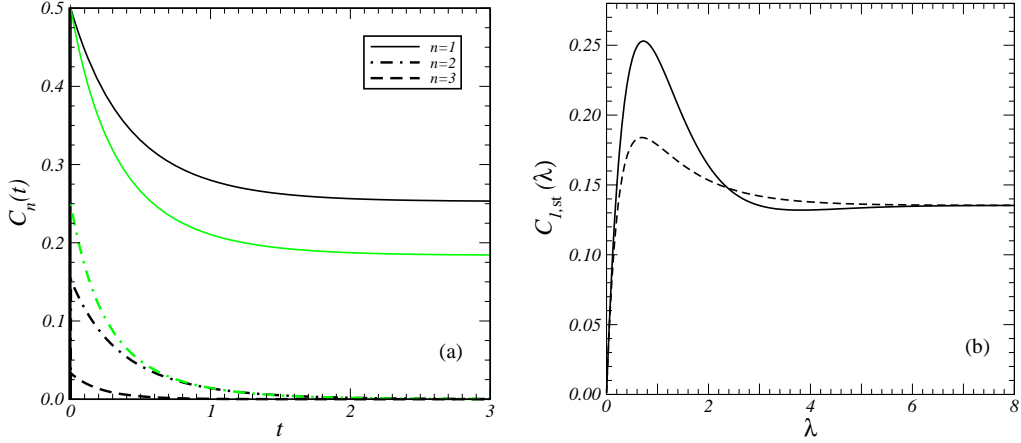

\centerline{\epsfxsize=6.5cm\epsfbox{kdh_fig9a.eps} ~~ \epsfxsize=6.5cm\epsfbox{kdh_fig9b.eps} }
\caption{\label{fig9} (a) Time-dependence of the probability $C_n(t)$ of at least $n$ consecutive occupied sites, for
$n=[1,2,3]$ from top to bottom and for an initial Poisson distribution with parameter $\lambda=1$. 
The grey lines give the corresponding results for an uncorrelated initial state with 
average density $\rho=0.5$. (b) Stationary particle-density 
$C_{1,\mbox{\rm\footnotesize st}}(\lambda)$ for the initial Poisson distribution(full line) and comparison
with the corresponding results for an initially uncorrelated state, with density $\rho\doteq 1-e^{-\lambda}$ (dashed line). 
}
\end{figure}

In order to illustrate what might happen for correlated initial conditions, we first observe that
the initial values $C_n(0)= \sum_{\ell=n}^{\infty} p_{\ell}$ of at least 
$n$ consecutive initial sites are rather cumulative probabilities
such that $p_n$ are the probabilities of having exactly $n$ 
consecutive occupied sites. As a simple example, consider the
case of an initial Poisson distribution $p_n = \lambda^n e^{-\lambda}/n!$ with parameter $\lambda$, such that
\BEQ
C_n(0) = \sum_{\ell =n}^{\infty} \frac{1}{\ell !} \lambda^{\ell} \,e^{-\lambda}
\EEQ
{}From eq.~(\ref{SolFinCorr}) one  immediately has
\BEQ
C_n(t) = \sum_{m=0}^{\infty} \frac{2^m (e^{-2t}-1)}{m!} 
\left( 1 - \frac{\Gamma(m+n,\lambda)}{\Gamma(m+n)}\right) \: e^{-2(n-1)t}
\EEQ
where $\Gamma(a,z)$ is an incomplete Gamma function \cite{Abra}. 
Although we did not succeed to convert this last sum into a known function, 
in figure~\ref{fig9}a we show the relaxation
for the first few values of $n$ and also compare with the uncorrelated case, 
already discussed above in  figure~\ref{fig1}. 
The comparison is possible since the parameters $\rho$ and $\lambda$ 
can be related by $C_1(0)=1-e^{-\lambda} \doteq \rho$. 
The stationary density is shown in figure~\ref{fig9}b and  displays, 
again, a non-monotonous dependence on the initial condition.  
This example illustrates that because of the large number of stationary
states, of the order $\sim 1.62^L$ on a lattice of $L$ sites \cite{Carl01,deSm02,Dutta09}, 
macroscopic quantities such as mean particle-densities depend on the
precise form of the initial condition and  their correlations. 

A physically motivated example for a correlated initial condition is 
the two-species irreversible pair annihilation
$A+B\to\emptyset+\emptyset$ of immobile particles $A$ and $B$. 
If $C_{1,A}(t)$ and $C_{1,B}(t)$ denote the average densities of the
particles $A$ and $B$, it can be shown \cite{Maju93} 
that $C_1(t)=\demi\left(C_{1,A}(t)+C_{1,B}(t)\right)$ 
is part of a set of quantities $C_n(2t)$ which
satisfy eq.~(\ref{eqEvolCn}), but with the initial conditions
\BEQ
C_n(0) = \left\{ \begin{array}{ll} \rho^n & \mbox{\rm ~~;~ $n$ pair} \\ 
                                   \demi(\rho_A+\rho_B) \rho^{n-1} & \mbox{\rm ~~;~ $n$ impair} 
                                   \end{array} \right. \;\; , \;\; \rho= \sqrt{\rho_A \rho_B\,} 
\EEQ
and where $\rho_{A,B}$ are the initial average densities of the $A,B$-particles, respectively. 
We shall return to this model in much detail in section~4.

\subsection{Correlation of two $n$-strings}

We search for exact correlators of a $n$-string with a $m$-string, 
separated by a certain distance $r$, which leads us to
compute the averages for observables such as 
$\underbrace{\bullet\bullet\ldots\bullet}_{n} \fbox{~r~} \underbrace{\bullet\bullet\ldots\bullet}_{m}$, 
where the state of the $r$ central sites is unknown. We shall also
refer to this as a `string with a hole of $r$ sites'.

\subsubsection{\underline{String with an one-site hole}}
In order to extend the methods of the previous subsection, 
we begin with the case of a single-site hole of
the form $\underbrace{\bullet\bullet\ldots\bullet}_{n} \fbox{~1~} 
\underbrace{\bullet\bullet\ldots\bullet}_{m}$ and the average 
   \be C_{n,m}^1(t) :=\langle \eta_1(t)\ldots\eta_{n}(t)
   \eta_{n+2}(t)\ldots\eta_{n+m+1}(t)\rangle
   \ee
For spatially translation-invariant initial conditions, the equation of motion is
    \be\fl {\D \over \D t}C_{n,m}^1(t)=-2\big[C_{n+1,m}^1(t)+C_{n,m+1}^1(t)
      +(n+m-2)C_{n,m}^1(t)+2C_{n+m+1}(t)\big] \label{24}
    \ee
In appendix~A, it is shown that the unique solution explicitly reads 
   \ba 
   C^1_{n,m}(t)&=\left(1-e^{4t}\right)C_{n+m+1}(t)\nonumber\\
   &\quad+\sum_{k,l=0}^{\infty}{C_{n+k,m+l}^1(0)\over k!\ \!l!}
       \left(e^{-2t}-1\right)^{k+l}e^{-2(n+m-2)t}
       \label{C:30}
   \ea
The correctness of this expression can be directly confirmed by checking the required initial conditions and by insertion into (\ref{24}).
The linear system of equations (\ref{24}) is known to have an unique solution, see e.g. \cite{Kamke59}.  

In the special case of a set of initially uncorrelated particles, with average density $\rho$ 
(or uncorrelated kinks in the {\sc kdh}-model), we have 
$C_{n,m}^{1}(0)=\rho^{n+m}$, $C_n(0)=\rho^n$ 
and the $(n,m)$-correlator for a hole of
size $r=1$ reads 
    \be \label{eq:Cnm1}
    C^1_{n,m}(t)=\rho^{n+m}\left[\rho\left(e^{-4t}-1\right)+1\right]
    e^{2\rho\left(e^{-2t}-1\right)-2(n+m-2)t}
    \ee
As expected, a finite value in the stationary limit $t\to\infty$ is only found for $n=m=1$. 

\subsubsection{\underline{String with a hole of $r$ sites}}
We now consider the $(n,m)$-correlator of two strings of particles (kinks) separated by a
hole consisting of $r>1$ sites and define the average 
   \be C_{n,m}^r(t) :=\langle \eta_1(t)\ldots\eta_{n}(t)
   \eta_{n+r+1}(t)\ldots\eta_{n+m+r}(t)\rangle\ee
Assuming translation-invariant initial conditions, the equation of motion is, see also \cite{deSm02}
    \be\fl{\D \over \D t}C_{n,m}^r(t)=-2\big[C^r_{n+1,m}(t)+C^r_{n,m+1}(t)
      +(n+m-2)C_{n,m}^r(t)+C_{n+1,m}^{r-1}(t)+C_{n,m+1}^{r-1}(t)\big]
    \ee
The exact and unique solution, for any given initial condition $C_{n,m}^{r}(0)$ is (see appendix~A)
\BEA
\hspace{-2.2truecm}C_{n,m}^{r}(t)&\hspace{-1.0truecm}=&\hspace{-0.4truecm}e^{-2(n+m-2)t} 
\sum_{k,l=0}^{\infty} \sum_{q=0}^{r-1} \sum_{q'=0}^{q} \left(\vekz{q}{q'}\right) 
C_{n+k+q-q',m+l+q'}^{r-q}(0) \frac{(e^{-2t}-1)^{k+l+q}}{k!\ l!\ q!} 
\nonumber \\
& & +2^r \left( \frac{(e^{-2t}-1)^r}{r!} + \frac{(e^{-2t}-1)^{r+1}}{(r+1)!} \right) 
e^{2(r+1)t} C_{n+m+r}(t) 
\label{53}
\EEA
where we also used the explicit expression (\ref{SolFinCorr}) 
for the $n$-string density $C_n(t)$. 
{\em This exact expression of the $(n,m)$-correlator 
$C_{n,m}^{r}(t)$ of two strings, separated by a distance $r$, 
for any initial condition, is the main result of this section.} 

\begin{figure}[tb]
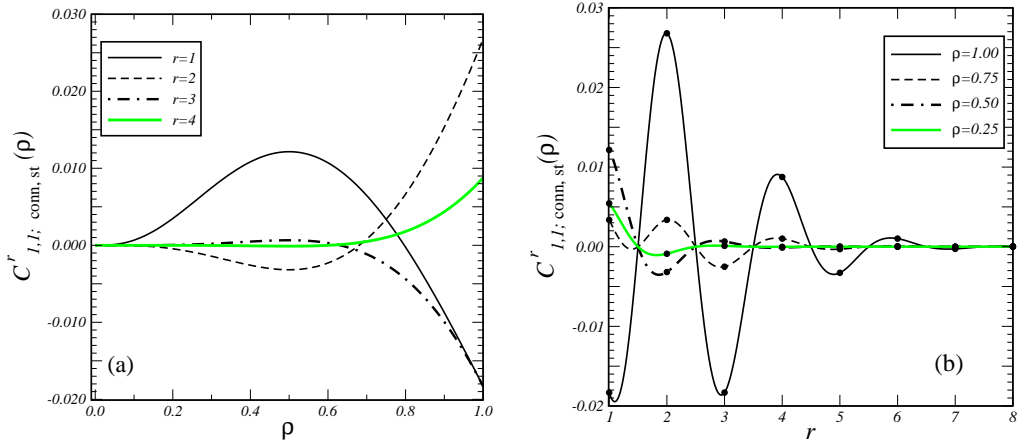

\centerline{\epsfxsize=6.5cm\epsfbox{kdh_fig3a.eps} ~~ \epsfxsize=6.5cm\epsfbox{kdh_fig3b.eps} }
\caption[fig3]{\label{fig3} Stationary connected particle-particle correlator 
$C_{1,1;\mbox{\rm conn., st}}^{r}$. 
(a) Dependence on the initial density $\rho$, for several values of the particle distance $r$. 
(b) Continuous interpolation of the dependence on the particle distance $r$, 
for several values of the density $\rho$. 
The full dots give the values for $r$ integer.  
}
\end{figure}

In the special case of initially uncorrelated particles with average density 
$\rho$, one has $C_{n,m}^{r}(0)=\rho^{n+m}$ so that
evaluation of the sums in (\ref{53}) leads to
\BEA
\lefteqn{C_{n,m}^{r}(t) - C_n(t) C_m(t) = \frac{\rho^{n+m}}{r!} e^{-2(n+m-2)t} \left( 2\rho\left( e^{-2t}-1\right)\right)^r }
\nonumber \\
& & \times \left[ \frac{r+e^{-2t}}{r+1} \exp\left(2\rho \left(e^{-2t}-1\right)\right) 
- {}_1F_{1}\left(r,r+1;2\rho\left(1-e^{-2t}\right)\right) \right]
\label{54}
\EEA
where ${}_1F_{1}$ is the confluent hyper-geometric function \cite{Abra}, 
in agreement with earlier results \cite{deSm02}. 
For a fixed time $t$, we observe a factorial decrease with $r$ of the connected correlator, which
is one of the main differences with what would have been predicted for weakly tapped granular matter
using Edwards' hypothesis which gives a more slow exponential decay \cite{Edwa94,deSm02,Godr05}. 
For $r\gg 1$ large, we find once more the double exponential time-dependence. 
For the first few values of $r$, we illustrate 
in figure~\ref{fig3} some aspects of the connected correlator 
$C_{n,m}^{r}(t)-C_n(t)C_m(t)$ in the stationary limit $t\to\infty$. 
In figure~\ref{fig3}a the dependence on $\rho$ is indicated 
for some small values of $r$, whereas in figure~\ref{fig3}b, the dependence
on the distance $r$ is continuously interpolated from (\ref{54}). 
Clearly, the non-trivial correlations decay very rapidly
with increasing $r$ such that in the stationary state, only very short-ranged correlations persist. 

\subsection{Some numerical tests}

In order to check our results further, we also also performed Monte Carlo simulations of the
pair-annihilation process $A+A\to\emptyset+\emptyset$ with immobile particles on a chain with
$L=65536$ sites. An average over $2\cdot 10^5$ histories was performed, in order to achieve a
statistical precision of the order $\sim 10^{-6}$. Three different initial conditions were
considered, namely
\BEQ \label{eq:estadosiniciais}
\begin{array}{ll}
\mbox{\rm a)} & \ldots AAAAAAAAAAA \ldots \\
\mbox{\rm b)} & \ldots AAA\emptyset AAA\emptyset AAA\emptyset AAA\emptyset\ldots \\
\mbox{\rm c)} & \ldots AAAA\emptyset AAAA\emptyset AAAA\emptyset AAAA\emptyset\ldots
\end{array}
\EEQ
The last two of these represent correlated initial states, 
with either three or four occupied sites followed by a single-site hole. 

{}From the results of the two previous sub-sections, both the density of the $n$-strings $C_n(t)$
as well as the $(n,m)$-correlators $C_{n,m}^r (t)$ can be found. 
\begin{figure}[tb]
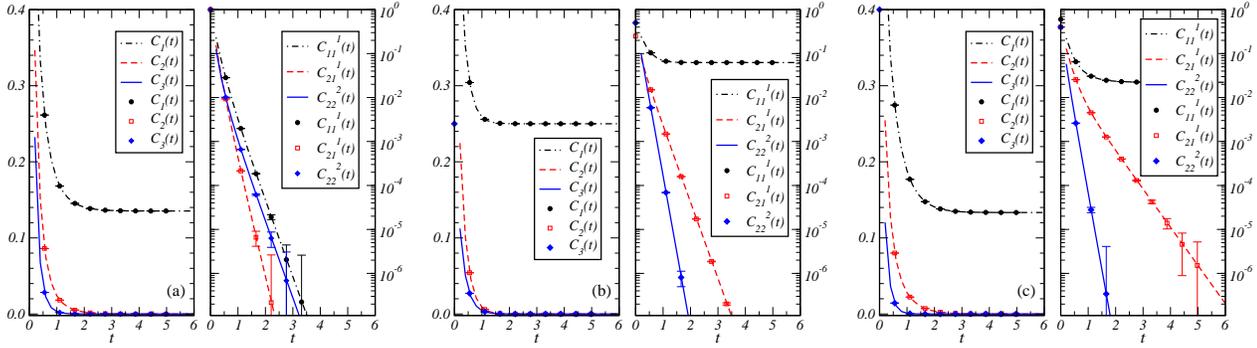

\centerline{\epsfxsize=5.3cm\epsfbox{kdh_fig10a.eps} ~ \epsfxsize=5.3cm\epsfbox{kdh_fig10b.eps} ~ \epsfxsize=5.3cm\epsfbox{kdh_fig10c.eps}}
\caption[fig10]{\label{fig10} Comparison of the predicted density $C_n(t)$ of the $n$-strings
(left panels) and of the correlators $C_{n,m}^r(t)$ (right panels) with Monte Carlo data for the
three initial conditions (a,b,c) as defined in eq.~(\ref{eq:estadosiniciais}), of the
pair-annihilation model on a chain with $L=65536$ sites. The lines are the predictions taken from 
(\ref{SolFinCorr},\ref{53}), where the time was rescaled by a factor $0.553$
throughout. The symbols give the Monte Carlo results.    
}
\end{figure}
The result of this comparison is shown in figure~\ref{fig10}abc, for the three initial conditions
defined in (\ref{eq:estadosiniciais}). We also checked, where applicable, that indeed 
$C_{n,m}^r(t)=C_{m,n}^r(t)$. It is necessary to re-scale the `Monte Carlo time-scale' by $0.553$ in order to match the implicit time scale in our equations. When this is done, one has from 
figure~\ref{fig10} a perfect agreement, both for densities and correlators, and for both uncorrelated
as well as uncorrelated initial states.\footnote{Eq.~(\ref{eq:Cnm1}) implies that $C_{11}^1(t)\to 0$
as $t\to\infty$ for an initially fully occupied lattice with $\rho=1$.}  

\subsection{Densities of $n$-strings on Bethe lattices}

\begin{figure}[tb]
\centerline{\epsfxsize=6.5cm\epsfbox{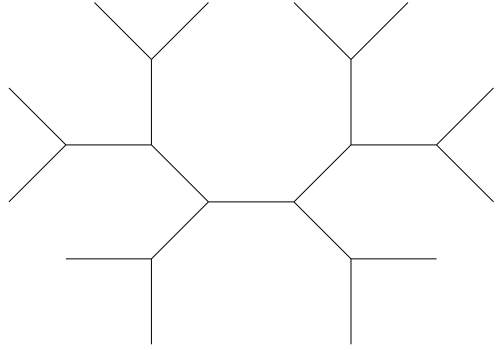} }
\caption{\label{fig_Bethe} A small portion of the Bethe lattice with $z=3$ neighbours.
}
\end{figure}

We now consider a Bethe lattice\footnote{Following \cite{Baxter82}, 
the term `{\em Bethe lattice}' denotes the deep interior
of the Cayley tree such that the strong surface effects of the latter do not arise for the Bethe lattice.} 
with $z$ links per node, see figure~\ref{fig_Bethe}, and we are
interested in $n$-string averages 
$C_n(t):=\langle \eta_i(t)\eta_j(t) \eta_k(t)
\ldots\rangle$ composed of a string of $n$ contiguous kinks.
The evolution equation for $C_n(t)$ involves three kinds of terms \cite{Maju93,Tome01}:
two terms $(z-1)C_{n+1}(t)$ due to the action of the transition rates on
the tail or the head of the string, $(n-1)$ terms $C_n(t)$
corresponding to the action of the transition rates on two
neighbouring kinks of the string and $(n-2)$ terms associated to
branching of the string. In the following, {\em we shall make the important
assumption that these last terms can be approximated by $C_{n+1}(t)$},
i.e. that the correlation depends only on the number of kinks and not
on their relative arrangement. This holds true if this condition is
realised by the initial conditions. The equation of motion then reduces to, for all $n\geq 1$
    \be \label{55} 
    {\D\over \D t}C_n(t)=-2(n-1)C_n(t) -2\left(n(z-2)+2\right) C_{n+1}(t)
    \ee
We see that the special case $z=2$ reduces to the case of the linear chain studied above. 
The unique solution is found by a slight generalisation of the methods used so far, as explained in appendix~C. 
The result is $\forall n\geq 1$
    \be \label{68} 
    \hspace{-1.0truecm}C_n(t)=
    \sum_{k=0}^{\infty} {(z-2)^k\over k!}{\Gamma\left(n+k+{2\over z-2}\right)
     \over\Gamma\left(n+{2\over z-2}\right)}\, C_{n+k}(0)\left(e^{-2t}-1\right)^k
    e^{-2(n-1)t}
    \ee
This is the main result of this subsection. 
Because of the asymptotic relation $x^{b-a} \Gamma(x+a)/\Gamma(x+b) = 1 +{\rm O}(1/x)$ \cite[Eq. (6.1.47)]{Abra}, in the
limit $z\to 2$ one recovers the result (\ref{Soluns}) or the linear chain. 

For the special case of initially uncorrelated particles such that $C_n(0)=\rho^n$, 
we reproduce the well-known form \cite{Evan84,Maju93,Abad04} 
\BEQ
C_n(t) = \rho^n e^{-2(n-1)t} \left[ 1 + (z-2)\rho\left( 1-e^{-2t}\right)\right]^{-(2+(z-2)n)/(z-2)}
\EEQ
which in the limit $z\to 2$ indeed reproduces (\ref{21}). 
We also see that for $z\geq 3$, the characteristic double-exponential
time-dependence becomes a simple exponential behaviour. 
This means that the double-exponential forms encountered above are a
peculiar feature of one-dimensional diffusion-free systems. In figure~\ref{fig1}b we illustrate
the dependence of the stationary particle-density $C_{1,\mbox{\rm st}}(\rho)$ on the initial density $\rho$,
for several values of $z$. Upon increasing $z$, this becomes increasingly like the mean-field result.

\section{Pair-annihilation model: the semi-infinite case}

We now extend the considerations of section~2 to the case when the initial conditions are no longer
required to be translation-invariant. At the end, we shall use this to derive densities and correlators
of $n$-strings for a semi-infinite system. 

\subsection{Densities of $n$-strings with inhomogeneous initial conditions}

When initial conditions are no longer invariant under translation, one
has to keep track of the first site of the string
    \be C_{a;n}(t) :=\langle \eta_a(t)\eta_{a+1}(t)
    \ldots\eta_{a+n-1}(t)\rangle\ee
where $n\in\mathbb{N}$ as before and $a\in\mathbb{Z}$ runs over all sites of the infinite chain. 
The evolution equation (\ref{eqev1}) reads
    \be {\D \over \D t}C_{a;n}(t)
    =-2(n-1)C_{a;n}(t)-2C_{a-1;n+1}(t)-2C_{a;n+1}(t)\ee
In appendix~A the unique solution of the desired $n$-string density is derived and reads 
   \be C_{a;n}(t)=\sum_{m,k=0}^{\infty}
   {C_{a-k;m+k+n}(0)\over m!\ \!k!}
   \left(e^{-2t}-1\right)^{m+k}\; e^{-2(n-1)t}
   \label{79}
   \ee
As seen above in section~2, only for $n=1$ a non-vanishing stationary density persists. 
The particular case of homogeneous initial conditions (\ref{SolFinCorr}) is recovered when
setting $C_{a-k;m+k+n}(0)=C_{m+k+n}(0)$ and rearranging indices.

\begin{figure}[tb]
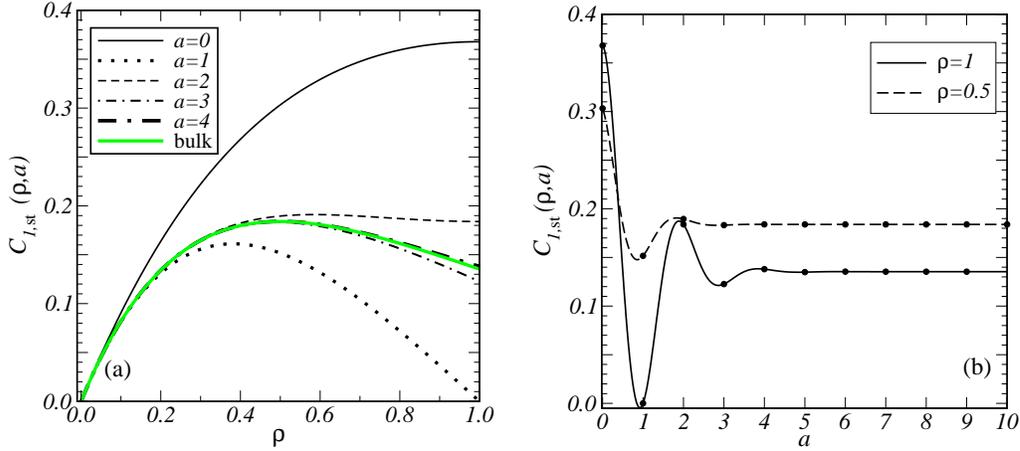

\centerline{\epsfxsize=6.5cm\epsfbox{kdh_fig2a.eps} ~~ \epsfxsize=6.5cm\epsfbox{kdh_fig2b.eps} }
\caption[fig2]{\label{fig2} (a) Stationary particle-density $C_{1,\mbox{\rm st}}(\rho,a)$ on a semi-infinite
chain, as a function of
the initial density $\rho$ for the distances $a=[0,2,4,3,1]$ from the boundary at $a=0$, from top to bottom 
(dash-dotted lines for $a\geq 1$). 
(b) Continuous interpolation of the dependencies of $C_{1,\mbox{\rm st}}(\rho,a)$ on $a$, for several
values of $\rho$. The full dots give the values of $C_{1,\mbox{\rm st}}(\rho,a)$ for $a$ integer. 
}
\end{figure}

We illustrate the physical content by considering the initial condition
\BEQ \label{79ini}
C_{a;n}(0) = \Theta(a) \rho^n \;\; , \;\; \Theta(a) = \left\{ 
\begin{array}{ll} 1 & \mbox{\rm ~~;~ if $a\geq 0$} \\ 
                  0 & \mbox{\rm ~~;~ if $a<0$} 
\end{array} \right. 
\EEQ
which describes a semi-infinite system confined to the positive half-axis 
$a\geq 0$ and with uncorrelated particles of
average density $\rho$. First, our result gives an interpolation between the particle density
$C_{0;1}(t) = \rho \exp\left(\rho \left(e^{-2t}-1\right)\right) 
\stackrel{t\to\infty}{\longrightarrow} \rho e^{-\rho}$ 
right at the surface and the form (\ref{21}) 
$C_{\infty;1}(t) = \rho \exp\left(2\rho\left( e^{-2t}-1\right)\right) 
\stackrel{t\to\infty}{\longrightarrow} \rho e^{-2\rho}$
deep in the bulk. Right at the surface, the stationary particle density varies monotonously with 
$\rho$ which is no longer the
case if one penetrates into the lattice, see figure~\ref{fig2}a. From (\ref{79}), 
we find for the stationary particle density
\BEA
\hspace{-1.0truecm}C_{1,\mbox{\rm\footnotesize st}}(\rho;a) &:=& 
\lim_{t\to\infty} C_1(t) = \rho e^{-\rho} \sum_{k=0}^a \frac{(-\rho)^k}{k!} 
\nonumber \\
&=& \rho e^{-2\rho} \left[ 1 + (-1)^a \rho^{1+a}\: {}_1F_{1}\left( a+1,a+2;\rho\right)/\Gamma(a+2) \right]
\EEA

In figure~\ref{fig2}, several aspects of this stationary density are displayed. 
Figure~\ref{fig2}a shows the dependence on $\rho$ for
several distances $a$ from the boundary. The asymptotic bulk form is rapidly reached, 
a few lattice sites away from the surface are sufficient. 
However, the approach towards to bulk is not monotonous. 
The approach towards the bulk becomes considerably slower with increasing
initial particle-density. Figure~\ref{fig2}b shows 
the dependence of the stationary density on the penetration depth $a$, for
a fixed value of $\rho$. We clearly observe an oscillatory behaviour as a function of $a$. 
This is natural, since in the stationary
state both nearest neighbours of an occupied site must be empty, 
which leads to an effective anti-correlation of the particles.

\subsection{Correlation of two $n$-strings with inhomogeneous initial conditions}

Analogously to section~2, it is useful to consider first a $(n,m)$-correlator of two strings which are
separated by a hole of a single site. Consider the averages
\BEQ
C_{a;n,m}^1(t) := \left\langle \eta_a \eta_{a+1}
\ldots \eta_{a+n} \eta_{a+n+2}\ldots \eta_{a+n+m} \right\rangle
\EEQ
which satisfies the equation of motion
\BEA
\hspace{-1.0truecm}\frac{\D}{\D t} C_{a;n,m}^1(t) &=& - 2(n+m-2) C_{a;n,m}^1(t) \nonumber \\
& & - 2 C_{a-1;n+1,m}^1(t) - 2 C_{a;n,m+1}^1(t) - 4C_{a;n+m+1}(t)
\label{C:45}
\EEA
which naturally generalises (\ref{24}). Similarly, for a hole with $r\geq 2$ sites, one has 
\BEQ
C_{a;n,m}^r(t) := \left\langle \eta_a \eta_{a+1}\ldots \eta_{a+n} \eta_{a+n+r+1}\ldots \eta_{a+n+m+r} 
\right\rangle
\EEQ
and which satisfies the equation of motion, with $r\geq 2$
\BEA
\hspace{-1.0truecm}\lefteqn{\frac{\D}{\D t} C_{a;n,m}^r(t) = - 2(n+m-2) C_{a;n,m}^r(t)} \nonumber \\
& & \hspace{-1.0truecm}- 2 C_{a-1;n+1,m}^r(t) - 2 C_{a;n,m+1}^r(t) 
    -  2 C_{a;n+1,m}^{r-1}(t) - 2 C_{a;n,m+1}^{r-1}(t)
\label{C:47}
\EEA
The solution of the equations (\ref{C:45},\ref{C:47}) is constructed in appendix~A, with the result, valid for all 
$n,m\in\mathbb{N}$, $a\in\mathbb{Z}$ and $r\geq 1$
\BEA
\hspace{-2.0truecm}\lefteqn{C_{a;n,m}^{r}(t) = 2^r 
\left( \frac{\left(e^{-2t}-1\right)^r}{r!} + 
\frac{\left(e^{-2t}-1\right)^{r+1}}{(r+1)!}\right)\: e^{2(r+1)t} C_{a;n+m+r}(t) } \nonumber \\
&&\hspace{-2.4truecm}+e^{-2(n+m-2)t} \sum_{\ell,k=0}^{\infty} 
\sum_{q=0}^{r-1} \sum_{q'=0}^{q} C_{a-\ell;n+\ell+q-q',m+k+q'}^{r-q}(0) \left(\vekz{q}{q'}\right) 
\frac{\left(e^{-2t}-1\right)^{k+\ell+q}}{k!\ \ell!\ q!}
\label{97}
\EEA
which is the second central result of this work. 

\begin{figure}[tb]
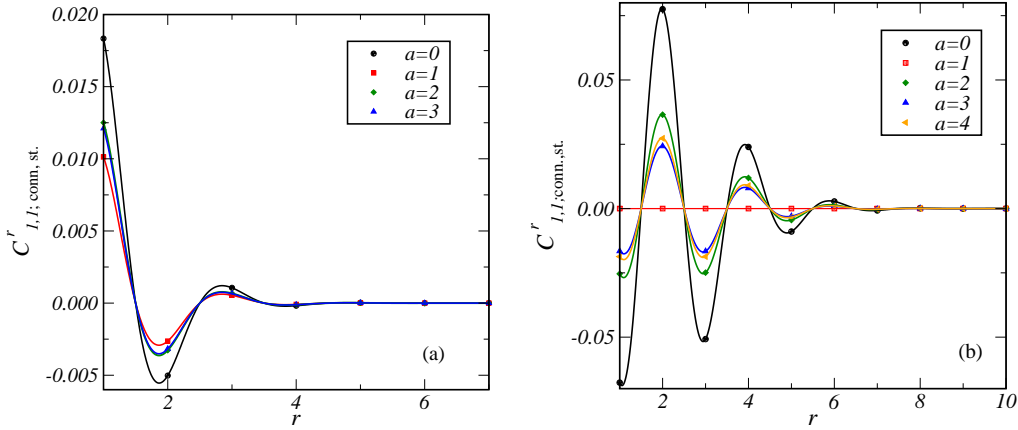

\centerline{\epsfxsize=6.5cm\epsfbox{kdh_fig7a.eps} ~~ \epsfxsize=6.5cm\epsfbox{kdh_fig7b.eps} }
\caption[fig7]{\label{fig7} Stationary connected particle-particle correlator 
$C_{1,1;\mbox{\rm conn., st}}^{r}$ from eq.~(\ref{97unc}), 
for the initial densities (a) $\rho=1/2$ and (b) $\rho=1$, 
and for several penetration depths $a$, counted from the boundary.   
The full symbols give the values for $r$ integer.  
}
\end{figure}

We illustrate the physical content of this result for initially uncorrelated particles 
of average density $\rho$, confined to the right half-plane. 
Then the initial correlator is $C_{a;n,m}^{r}(0)=\rho^{n+m}\Theta(a)$ and the initial density
was given in (\ref{79ini}). The stationary connected particle-particle correlator is read off from (\ref{97}), 
where $\Gamma(a,z)$ is an incomplete Gamma function \cite{Abra}, 
\BEA
\hspace{-2.0truecm}\lefteqn{ C_{a;1,1}^{r} - C_{a;1}(\infty) C_{a+r+1;1}(\infty) }  
\label{97unc} \\
\hspace{-2.0truecm}&=& \rho^2 e^{-2\rho} \frac{\Gamma(1+a,-\rho)}{\Gamma(1+a)} \left[
\frac{e^{-\rho} \Gamma(1+r,-2\rho)}{\Gamma(1+r)} - \frac{e^{-2\rho} \Gamma(1+r+a,-\rho)}{\Gamma(1+r+a)} 
 - \frac{(-2\rho)^r}{\Gamma(r+2)} \right]
 \nonumber
\EEA
and depends not only on the distance $r$ between the particles but also on the penetration depth $a$
of the leftmost one. This connected correlator therefore describes correlations between a particle close
to the surface and another one more deeply situated in the bulk. 
In figure~\ref{fig7}, the connected stationary particle-particle correlator and its
factorial spatial decay is displayed for two values of
the particle-density. It shape strongly evolves with changes of the penetration depth $a$.  
 
\section{Two-species annihilation model without diffusion}

We now consider {\em two} species of immobile particles, denoted by $A$ and $B$ that can occupy
the nodes of an infinite chain. The only admissible reaction is
the pair-annihilation between nearest-neighbour particles:
$A+B\rightarrow\emptyset+\emptyset$, with rate $1$. As in the single-species model treated above, the
number of stationary states on a chain of $L$ sites grows exponentially with $L$ and is of the
order $\sim\left(\sqrt{2\,}+1\right)^L$, see eqs.~(\ref{A5},\ref{A7}) in 
the appendix~D for details. If $\eta_i\in\{\emptyset,A,B\}$ denotes the
occupation of the $i$-th site, the transition rates in the master equation (\ref{1}) read
   \be\fl \omega_i(\{\eta\}\rightarrow\{\eta'\})
   =\sum_i \left[\delta_{\eta_i,A}\delta_{\eta_{i+1},B}
     +\delta_{\eta_i,B}\delta_{\eta_{i+1},A}\right]
   \delta_{\eta_i',\emptyset}\delta_{\eta_{i+1}',\emptyset}
   \prod_{j\not\in\{i,i+1\}} \delta_{\eta_j,\eta_j'}\ee

\subsection{Probability of an alternating $n$-string $ABABAB\ldots$}

\subsubsection{General solution}
Since reactions occur whenever $AB$-pairs meet, we consider the probability of having sequences of 
$n$ particles, 
consisting of uninterrupted pairs $ABABAB\ldots$ or $BABABA\ldots$
   \ba A_n :=\Pr\big(\underbrace{ABABAB\ldots}_{n\; \mbox{\rm\footnotesize 
   sites}}\big)=\langle\delta_{\eta_1,A}\delta_{\eta_2,B}
   \delta_{\eta_3,A}\delta_{\eta_4,B}\ldots\rangle\nonumber\\
   B_n :=\Pr\big(\underbrace{BABABA\ldots}_{n\; \mbox{\rm\footnotesize 
   sites}}\big)=\langle\delta_{\eta_1,B}\delta_{\eta_2,A}
   \delta_{\eta_3,B}\delta_{\eta_4,A}\ldots\rangle
   \ea
Starting from the master equation
   \be\fl{\D\over\D t}\Pr(\{\eta\};t)
   =\sum_{\{\eta'\}}
   \left[\Pr(\{\eta'\};t)\omega_i(\{\eta'\}\rightarrow\{\eta\})
     -\Pr(\{\eta\};t)\omega_i(\{\eta\}\rightarrow\{\eta'\})\right]\ee
the evolution equation of $A_n$ is
   \ba\fl{\D\over\D t}A_n=\sum_{\{\eta\},\{\eta'\}}
   \delta_{\eta_1,A}\delta_{\eta_2,B}\ldots
   \left[\Pr(\{\eta'\};t)\omega_i(\{\eta'\}\rightarrow\{\eta\})
     -\Pr(\{\eta\};t)\omega_i(\{\eta\}\rightarrow\{\eta'\})\right]
   \nonumber\\
   \lo=-\sum_{\{\eta\},\{\eta'\}}\left(\delta_{\eta_1,A}\delta_{\eta_2,B}\ldots
   -\delta_{\eta_1',A}\delta_{\eta_2',B}\ldots\right)\omega_i(\{\eta\}\rightarrow\{\eta'\})
   \Pr(\{\eta\};t)\nonumber\\
   \lo=-\sum_{\{\eta\},\{\eta'\}}\sum_i\left(\delta_{\eta_1,A}\delta_{\eta_2,B}\ldots
   -\delta_{\eta_1',A}\delta_{\eta_2',B}\ldots\right)
   \Big[\prod_{j\not\in\{i,i+1\}} \delta_{\eta_j,\eta_j'}\Big]\nonumber\\
   \lo\quad\quad\quad\quad\times
   \left(\delta_{\eta_i,A}\delta_{\eta_{i+1},B}
     +\delta_{\eta_i,B}\delta_{\eta_{i+1},A}\right)
   \delta_{\eta_i',\emptyset}\delta_{\eta_{i+1}',\emptyset}\Pr(\{\eta\};t)
   \label{EqEvolGenAB}
   \ea
The two terms in the parenthesis cancel unless $j=i$ or $j=i+1$. When
it is the case, the second term vanishes because of the factor
$\delta_{\eta_i',\emptyset}\delta_{\eta_{i+1}',\emptyset}$. It remains
   \ba
   {\D\over\D t}A_n&=-\sum_{\{\eta\}}\sum_{i=0}^{n+1}
   \delta_{\eta_1,A}\delta_{\eta_2,B}\ldots
   \left(\delta_{\eta_i,A}\delta_{\eta_{i+1},B}
     +\delta_{\eta_i,B}\delta_{\eta_{i+1},A}\right)\Pr(\{\eta\};t)
     \nonumber\\
     &=\sum_{i=0}^{n+1} \langle\delta_{\eta_1,A}\delta_{\eta_2,B}\ldots
   \left(\delta_{\eta_i,A}\delta_{\eta_{i+1},B}
     +\delta_{\eta_i,B}\delta_{\eta_{i+1},A}\right)\rangle\nonumber\\
     &=-B_{n+1}-(n-1)A_n-A_{n+1}
     \label{ab1}
   \ea
where the first term is due to a transition rate acting on the first
site of the string (meaning that a $B$ was present in front), the
second to transitions in the bulk and the last to an annihilation
involving only the last site. Analogously, $B_n$ satisfies the
evolution equation
   \be \label{ab2}
   {\D\over\D t}B_n=-A_{n+1}-(n-1)B_n-B_{n+1}
   \ee
We obtain a set of two coupled differential equations similar to
(\ref{eqEvolCn}). The method of solution closely follows the one used in the previous sections and
is outlined in appendix~B. 
We find for the probabilities of the two $n$-strings
   \ba
   \fl A_n(t)={1\over 2}\left[A_n(0)-B_n(0)
   +\sum_{m=0}^{\infty} {2^m\over m!}\left(A_{n+m}(0)+B_{n+m}(0)\right)
   \left(e^{-t}-1\right)^m\right]e^{-(n-1)t}\nonumber\\
   \fl B_n(t)={1\over 2}\left[B_n(0)-A_n(0)
     +\sum_{m=0}^{\infty} {2^m\over m!}\left(A_{n+m}(0)+B_{n+m}(0)\right)
   \left(e^{-t}-1\right)^m\right]e^{-(n-1)t} \nonumber \\
   \fl ~ \label{109}
   \ea
This is the first main result of this section. 

\subsubsection{Homogeneous initial conditions}
We assume that $A$ and $B$ particles are initially randomly distributed on the
lattice without any correlation among them. If the densities are
denoted $\rho_A$ and $\rho_B$ (such that $1-\rho_A-\rho_B$ is the initial density of vacant sites) 
then
    \ba\fl A_n(0)=\rho_A^{[(n+1)/2]}\rho_B^{[n/2]}
    =(\rho_A\rho_B)^{n/2}\left[{1+(-1)^n\over 2}
      +{1-(-1)^n\over 2}\sqrt{\rho_A\over\rho_B}\right]
    \nonumber\\
    \fl B_n(0)=\rho_B^{[(n+1)/2]}\rho_A^{[n/2]}
    =(\rho_A\rho_B)^{n/2}\left[{1+(-1)^n\over 2}
      +{1-(-1)^n\over 2}\sqrt{\rho_B\over\rho_A}\right]
    \ea
where $[x]$ is the largest integer smaller than $x$. The densities of the $n$-strings follow from (\ref{109}) 

\begin{figure}[tb]
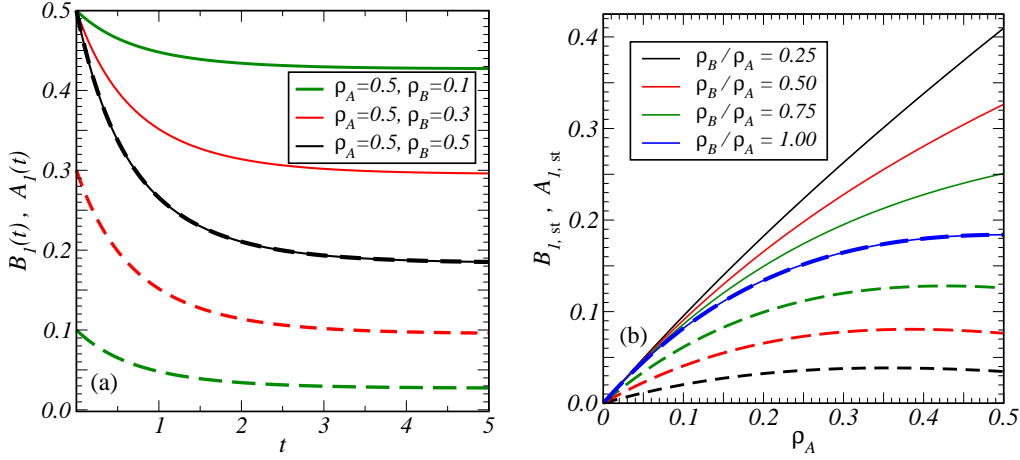

\centerline{\epsfxsize=6.5cm\epsfbox{kdh_fig4a.eps} ~~ \epsfxsize=6.5cm\epsfbox{kdh_fig4b.eps} }
\caption[fig4]{\label{fig4} (a) Time-dependence of the particle-concentrations $a(t)=A_1(t)$ (full lines), 
with $\alpha=0.5$ and
$\beta=[0.1,0.3,0.5]$ from top to bottom and $b(t)=B_1(t)$ (dashed lines) with $\alpha=0.5$ and
$\beta=[0.1,0.3,0.5]$ from bottom to top. (b) Stationary particle-densities 
$a(\infty)=A_{1,\mbox{\rm\footnotesize st}}$ (full lines)
and $b(\infty)=B_{1,\mbox{\rm\footnotesize st}}$ (dashed lines) as a function of the initial density $\rho_A$ 
and for
the ratios $\rho_B/\rho_A=[0.25,0.50,0.75,1.00]$ from top to bottom for the $A$-particles and from bottom to 
top for the $B$-particles. 
}
\end{figure}

\BEA
\hspace{-2.0truecm}A_n(t) &\hspace{-1.0truecm}=& \hspace{-0.5truecm}\demi (\rho_A \rho_B)^{n/2} e^{-(n-1)t} 
\left\{ \left[ \left( 1 + \demi\frac{\rho_A+\rho_B}{\sqrt{\rho_A \rho_B}} \right)
e^{2\sqrt{\rho_A\rho_B\,}\,\left( e^{-t}-1\right)} + \demi\frac{\rho_A-\rho_B}{\sqrt{\rho_A \rho_B}} \right] 
\right.\nonumber \\
&& \left.+ (-1)^n \left[ \left( 1 - \demi\frac{\rho_A+\rho_B}{\sqrt{\rho_A \rho_B}} \right)
e^{-2\sqrt{\rho_A\rho_B\,}\,\left( e^{-t}-1\right)} - \demi\frac{\rho_A-\rho_B}{\sqrt{\rho_A \rho_B}} \right] 
\right\}
\nonumber \\
\hspace{-2.0truecm}B_n(t) &\hspace{-1.0truecm}=& \hspace{-0.5truecm}\demi (\rho_A \rho_B)^{n/2} e^{-(n-1)t} 
\left\{ \left[ \left( 1 + \demi\frac{\rho_A+\rho_B}{\sqrt{\rho_A \rho_B}} \right)
e^{2\sqrt{\rho_A\rho_B\,}\,\left( e^{-t}-1\right)} - \demi\frac{\rho_A-\rho_B}{\sqrt{\rho_A \rho_B}} \right] 
\right.\nonumber \\
&& \left.+ (-1)^n \left[ \left( 1 - \demi\frac{\rho_A+\rho_B}{\sqrt{\rho_A \rho_B}} \right)
e^{-2\sqrt{\rho_A\rho_B\,}\,\left( e^{-t}-1\right)} + \demi\frac{\rho_A-\rho_B}{\sqrt{\rho_A \rho_B}} \right] 
\right\} 
\EEA
and which for $n=1$ reproduces the known result \cite{Maju93}. 

In figure~\ref{fig4}a, the relaxation of the particle-densities $A_1(t)$ and $B_1(t)$ 
towards their stationary values is illustrated. 
In figure~\ref{fig4}b, the stationary densities are shown themselves, 
as a function of the initial parameters $\rho_{A,B}$. 

\subsection{Correlations between two strings $ABAB\ldots$ separated by one site}
We now consider the probability to observe two strings separated by a
single-site hole
\newpage 
\typeout{*** ici saut de page *** }
   \BEA 
   \hspace{-2.0truecm}A^1_{n,m}&\hspace{-1.2truecm}:=& 
   \hspace{-0.0truecm}\Pr\big( \underbrace{ABABAB\ldots}_{n \mbox{\rm\footnotesize ~sites}} \fbox{~1~} 
   \underbrace{\ldots ABABAB\ldots}_{m \mbox{\rm\footnotesize ~sites}}\big)\nonumber\\
   &\hspace{-1.2truecm}=& \left\{ \begin{array}{ll} 
   \langle\delta_{\eta_1,A}\delta_{\eta_2,B}\ldots\delta_{\eta_n,A}\:
   \delta_{\eta_{n+2},A}\delta_{\eta_{n+3},B}\delta_{\eta_{n+4},A}\ldots\rangle & \mbox{\rm ~~;~ if $n$ 
   impair} \\
   \langle\delta_{\eta_1,A}\delta_{\eta_2,B}\ldots\delta_{\eta_n,B}\:
   \delta_{\eta_{n+2},B}\delta_{\eta_{n+3},A}\delta_{\eta_{n+4},B}\ldots\rangle & \mbox{\rm ~~;~ if $n$ pair} 
   \end{array} \right. 
   \EEA
By definition, the leftmost particle is $A$ (thus the name of this quantity).
The first block consists of $n$ particles of alternating species $A$ and
$B$. The second block similarly consists of $m$ particles of alternating types.
The important point is that the last particle of the first block is
the same as the first of the second one. The particles on even (odd)
sites are all of the same type. The state of the site in the hole is unknown. In the same manner, we also 
define the quantity
   \BEA 
   \hspace{-2.0truecm}B^1_{n,m}&\hspace{-1.2truecm}:=& 
   \hspace{-0.0truecm}\Pr\big( \underbrace{BABABA\ldots}_{n \mbox{\rm\footnotesize ~sites}} \fbox{~1~} 
   \underbrace{\ldots BABABA\ldots}_{m \mbox{\rm\footnotesize ~sites}}\big)\nonumber\\
   &\hspace{-1.2truecm}=& \left\{ \begin{array}{ll} 
   \langle\delta_{\eta_1,B}\delta_{\eta_2,A}\ldots\delta_{\eta_n,B}\:
   \delta_{\eta_{n+2},B}\delta_{\eta_{n+3},A}\delta_{\eta_{n+4},B}\ldots\rangle & \mbox{\rm ~~;~ if $n$ 
   impair} \\
   \langle\delta_{\eta_1,B}\delta_{\eta_2,A}\ldots\delta_{\eta_n,A}\:
   \delta_{\eta_{n+2},A}\delta_{\eta_{n+3},B}\delta_{\eta_{n+4},A}\ldots\rangle & \mbox{\rm ~~;~ if $n$ pair} 
   \end{array} \right. 
   \EEA
but here, the leftmost particle is $B$. By
exchanging $A$ and $B$ particles, $A^1_{n,m}$ is transformed into $B^1_{n,m}$.
The equations of motion are obtained using (\ref{EqEvolGenAB})
   \ba\fl&{\D\over\D t}A^1_{n,m}=-B^1_{n+1,m}-(n-1)A^1_{n,m}-2A_{n+m+1}
   -(m-1)A^1_{n,m}-A^1_{n,m+1}\nonumber\\
   \fl&{\D\over\D t}B^1_{n,m}=-A^1_{n+1,m}-(n-1)B^1_{n,m}-2B_{n+m+1}
   -(m-1)B^1_{n,m}-B^1_{n,m+1}\ea
where $A_n$ and $B_n$ are the correlations of a string of particles
without holes previously calculated. The solution of these equations is given in appendix~B and we find
{\small
   \ba\fl
   A^1_{n,m}(t)={1\over 2}\left[\big(e^{-t}-1\big)\big(e^{-t}+1\big)
     \left(A_{n+m+1}(0)-B_{n+m+1}(0)+\sum_{k=0}^{\infty}\left[A_{n+m+k+1}+B_{n+m+k+1}(0)\right]
          {2^k\big(e^{-t}-1\big)^k\over k!}\right)\right.\nonumber\\
   \fl\quad +\left.\sum_{k,l=0}^{\infty}\left((-1)^k\left[A^1_{n+k,m+l}(0)-B^1_{n+k,m+l}(0)\right]
   +\left[A^1_{n+k,m+l}(0)+B^1_{n+k,m+l}(0)\right]\right){\big(e^{-t}-1\big)^{k+l}\over k!\ \!l!}
   \right]e^{-(n+m-2)t}\ea
   }
and
{\small
   \ba\fl
   B^1_{n,m}(t)={1\over 2}\left[\big(e^{-t}-1\big)\big(e^{-t}+1\big)
     \left(B_{n+m+1}(0)-A_{n+m+1}(0)+\sum_{k=0}^{\infty}\left[A_{n+m+k+1}+B_{n+m+k+1}(0)\right]
          {2^k\big(e^{-t}-1\big)^k\over k!}\right)\right.\nonumber\\
   \fl\quad +\left.\sum_{k,l=0}^{\infty}\left((-1)^k\left[B^1_{n+k,m+l}(0)-A^1_{n+k,m+l}(0)\right]
   +\left[A^1_{n+k,m+l}(0)+B^1_{n+k,m+l}(0)\right]\right){\big(e^{-t}-1\big)^{k+l}\over k!\ \!l!}
   \right]e^{-(n+m-2)t}\ea
   }

\subsection{Correlations between two strings $ABAB\ldots$ separated by $r$ sites}

We now consider the probability to observe two strings separated by a
hole of $r\geq 2$ sites and define
   \BEA 
   \hspace{-2.0truecm}A^r_{n,m}&\hspace{-1.2truecm}:=& 
   \hspace{-0.0truecm}\Pr\big( \underbrace{ABABAB\ldots}_{n \mbox{\rm\footnotesize ~sites}} \fbox{~r~} 
   \underbrace{\ldots ABABAB\ldots}_{m \mbox{\rm\footnotesize ~sites}}\big) \\
   &\hspace{-2.3truecm}=& \hspace{-1.2truecm}\left\{ \begin{array}{ll} 
   \langle\delta_{\eta_1,A}\delta_{\eta_2,B}\ldots\delta_{\eta_n,A}\:
   \delta_{\eta_{n+r+1},A}\delta_{\eta_{n+r+2},B}\delta_{\eta_{n+r+3},A}\ldots\rangle & \mbox{\rm ~;~ if $n$ 
   impair, $r$ impair} \\
   \langle\delta_{\eta_1,A}\delta_{\eta_2,B}\ldots\delta_{\eta_n,B}\:
   \delta_{\eta_{n+r+1},B}\delta_{\eta_{n+r+2},A}\delta_{\eta_{n+r+3},B}\ldots\rangle & \mbox{\rm ~;~ if $n$ 
   pair, $r$ impair} \\
   \langle\delta_{\eta_1,A}\delta_{\eta_2,B}\ldots\delta_{\eta_n,A}\:
   \delta_{\eta_{n+r+1},B}\delta_{\eta_{n+r+2},A}\delta_{\eta_{n+r+3},B}\ldots\rangle & \mbox{\rm ~;~ if $n$ 
   impair, $r$ pair} \\
   \langle\delta_{\eta_1,A}\delta_{\eta_2,B}\ldots\delta_{\eta_n,B}\:
   \delta_{\eta_{n+r+1},A}\delta_{\eta_{n+r+2},B}\delta_{\eta_{n+r+3},A}\ldots\rangle & \mbox{\rm ~;~ if $n$ 
   pair, $r$ pair} 
   \end{array} \right. \nonumber
   \EEA
The leftmost particle is again $A$ (thus the name of this quantity).
The first block consists in $n$ particles of alternating types $A$ and
$B$. The hole corresponds to $r$ sites. The second block consists in
$m$ particles of alternating types. The important point is that the
particle types in the second block is completely determined by the
first block. If $n+r$ is odd, the first particle of the second block
is $B$ while it is $A$ if $n+r$ is even. The state of the $r$ sites in the hole in unknown. 
In the same way, we also define the quantity
   \BEA 
   \hspace{-2.0truecm}B^r_{n,m}&\hspace{-1.2truecm}:=& 
   \hspace{-0.0truecm}\Pr\big( \underbrace{BABABA\ldots}_{n \mbox{\rm\footnotesize ~sites}} \fbox{~r~} 
   \underbrace{\ldots BABABA\ldots}_{m \mbox{\rm\footnotesize ~sites}}\big) \\
   &\hspace{-2.3truecm}=& \hspace{-1.2truecm}\left\{ \begin{array}{ll} 
   \langle\delta_{\eta_1,B}\delta_{\eta_2,A}\ldots\delta_{\eta_n,B}\:
   \delta_{\eta_{n+r+1},B}\delta_{\eta_{n+r+2},A}\delta_{\eta_{n+r+3},B}\ldots\rangle & \mbox{\rm ~;~ if $n$ 
   impair, $r$ impair} \\
   \langle\delta_{\eta_1,B}\delta_{\eta_2,A}\ldots\delta_{\eta_n,A}\:
   \delta_{\eta_{n+r+1},A}\delta_{\eta_{n+r+2},B}\delta_{\eta_{n+r+3},A}\ldots\rangle & \mbox{\rm ~;~ if $n$ 
   pair, $r$ impair} \\
   \langle\delta_{\eta_1,B}\delta_{\eta_2,A}\ldots\delta_{\eta_n,B}\:
   \delta_{\eta_{n+r+1},A}\delta_{\eta_{n+r+2},B}\delta_{\eta_{n+r+3},A}\ldots\rangle & \mbox{\rm ~;~ if $n$ 
   impair, $r$ pair} \\
   \langle\delta_{\eta_1,A}\delta_{\eta_2,B}\ldots\delta_{\eta_n,B}\:
   \delta_{\eta_{n+r+1},B}\delta_{\eta_{n+r+2},A}\delta_{\eta_{n+r+3},B}\ldots\rangle & \mbox{\rm ~;~ if $n$ 
   pair, $r$ pair} 
   \end{array} \right. \nonumber
   \EEA
In spite of these complex-looking definitions, the equations of motions have the relatively simple form, 
found by using (\ref{EqEvolGenAB})
   \ba\fl
   &{\D\over\D t}A^r_{n,m}=-B^r_{n+1,m}-(n-1)A^r_{n,m}-A^{r-1}_{n+1,m}
     -A^{r-1}_{n,m+1}-(m-1)A^r_{n,m}-A^r_{n,m+1}\nonumber\\
   \fl &{\D\over\D t}B^r_{n,m}=-A^r_{n+1,m}-(n-1)B^r_{n,m}-B^{r-1}_{n+1,m}
   -B^{r-1}_{n,m+1}-(m-1)B^r_{n,m}-B^r_{n,m+1}\nonumber \\
   \fl & ~
   \ea
Following the same lines as before (see appendix~B for the details), the solution of these equation is, 
for all integers $r\geq 1$
  \ba\fl A^r_{n,m}(t)={1\over 2}\left[
     2^r\left({(e^{-t}-1)^{r+1}\over (r+1)!}+{(e^{-t}-1)^r\over r!}\right)
     \right. 
     \label{166} \\
     \hspace{-2.3truecm}\times\Big[ \left( A_{n+m+r}(0)-B_{n+m+r}(0)\right)
     +\sum_{c=0}^{\infty} \left(A_{n+m+c+r}(0)+B_{n+m+c+r}(0)\right){2^c(e^{-t}-1)^c\over c!}
       \Big]\nonumber\\
     \hspace{-2.3truecm}+\sum_{a,b=0}^{\infty}\sum_{l=0}^{r-1}\sum_{p=0}^l
       \big[\big(1+(-1)^{a}\big)A^{r-l}_{n+p+a,m+l-p+b}(0)\big.\nonumber\\
       \left.\big.+\big(1-(-1)^{a}\big)B^{r-l}_{n+p+a,m+l-p+b}(0)\big]
       {(e^{-t}-1)^{l+a+b}\over p!\ \!(l-p)!\ \!a!\ \!b!}\right]e^{-(n+m-2)t}\nonumber
\ea
and
  \ba\fl B^r_{n,m}(t)={1\over 2}\left[
     2^r\left({(e^{-t}-1)^{r+1}\over (r+1)!}+{(e^{-t}-1)^r\over r!}\right)
     \right.
     \label{167} \\
     \hspace{-2.3truecm}\times\Big[B_{n+m+r}(0)-A_{n+m+r}(0)+\sum_{c=0}^{\infty} 
     \left(A_{n+m+c+r}(0)+B_{n+m+c+r}(0)\right)
               {2^c(e^{-t}-1)^c\over c!}\Big]\nonumber\\
     \hspace{-2.3truecm}+\sum_{a,b=0}^{\infty}\sum_{l=0}^{r-1}\sum_{p=0}^l 
       \big[\big(1-(-1)^{a}\big)A^{r-l}_{n+p+a,m+l-p+b}(0)\big.\nonumber\\
       \left.\big.+\big(1+(-1)^{a}\big)B^{r-l}_{n+p+a,m+l-p+b}(0)\big]
       {(e^{-t}-1)^{l+a+b}\over p!\ \!(l-p)!\ \!a!\ \!b!}\right]e^{-(n+m-2)t}\nonumber
     \ea
which is our last main result. In the stationary state, only the density-density
correlations $A^r_{1,1}$ and $B^r_{1,1}$ survives.

\subsubsection{Homogeneous initial conditions}
\begin{figure}[tb]
\centerline{\epsfxsize=8cm\epsfbox{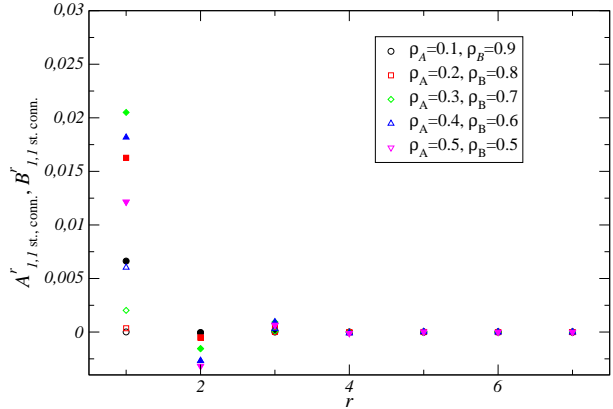}}
\caption[fig3]{\label{fig8} Stationary connected particle-particle correlator 
$A_{1,1;\mbox{\rm conn., st}}^{r}$ (open symbols) and $B_{1,1;\mbox{\rm conn., st}}^{r}$ (full symbols)
with respect to the distance $r$ between the two particles. The different symbols
correspond to different values of the initial densities.
}
\end{figure}

As before, we assume that $A$ and $B$ particles are initially randomly distributed on the
lattice without any correlation among them. The densities being denoted $\rho_A$ and $\rho_B$,
the initial correlations reads
    \ba\fl A^r_{nm}(0)=(\rho_A\rho_B)^{(n+m)/2}\left[{1+(-1)^n\over 2}
      +{1-(-1)^n\over 2}\sqrt{\rho_A\over\rho_B}\,\right]\\
    \lo\times \left[{1+(-1)^m\over 2}+{1-(-1)^m\over 2}\left(
        {1+(-1)^{n+r}\over 2}\sqrt{\rho_A\over\rho_B}
        +{1-(-1)^{n+r}\over 2}\sqrt{\rho_B\over\rho_A}\,\right)\right]\nonumber\\
    \fl B^r_{nm}(0)=(\rho_A\rho_B)^{(n+m)/2}\left[{1+(-1)^n\over 2}
      +{1-(-1)^n\over 2}\sqrt{\rho_B\over\rho_A}\,\right]\nonumber\\
    \lo\times \left[{1+(-1)^m\over 2}+{1-(-1)^m\over 2}\left(
        {1+(-1)^{n+r}\over 2}\sqrt{\rho_B\over\rho_A}
        +{1-(-1)^{n+r}\over 2}\sqrt{\rho_A\over\rho_B}\,\right)\right]\nonumber
    \ea
The connected density-density correlations in the stationary state, as given
by (\ref{166}) and (\ref{167}) in the limit $t\rightarrow +\infty$, are plotted
in figure \ref{fig8}. 
Qualitatively, the behaviour is similar to the one found above in the single-species model. 

\section{Conclusions}

We have been studying densities and correlators in 
interacting-particle models with immobile particles and irreversible
reactions which have an exponentially large number of stationary states. 
These models are relevant in the
context of {\sc rsa} or else for weakly tapped granular systems, since they have a non-vanishing 
configurational entropy (or `complexity'). 
The main results are as follows. In the case of a single type of particle with the annihilation reaction
$A+A\rar \emptyset+\emptyset$, we have computed exactly:
\begin{enumerate}
\item the density (\ref{SolFinCorr}) of the $n$-strings on the infinite chain
\item the string-string correlator (\ref{53}) of two $n$-strings on the infinite chain
\item the density (\ref{68}) of the 
$n$-strings on the Bethe lattice with arbitrary coordination number $z\geq 2$
\item the density (\ref{79}) of the 
$n$-strings without translation-invariance, including the semi-infinite chain
\item the string-string correlator (\ref{97}) of two 
$n$-strings without translation-invariance, including the semi-infinite chain
\end{enumerate}
We did check in all cases that the stated solutions indeed solve the respective equations of motion. 

In the case of two species of particles $A$ and $B$ evolving through the reaction 
$A+B\rar \emptyset+\emptyset$, we obtained the
\begin{enumerate}
\item density (\ref{109}) of alternating $AB$-strings of total length $n$,
\item the string-string correlator (\ref{166},\ref{167})
\end{enumerate}
All these results are given for {\em arbitrary} physically admissible initial configurations. 
Setting $n=1$, they include explicit expressions for the average particle-densities and their
correlators. 
In the special situation of initially uncorrelated particles 
of prescribed densities, we reproduce all previously obtained results 
\cite{Kemk81,Evan84,Evan93,Maju93,Oliv92,Oliv94,Prad01,Tome01,deSm02,Abad04} as special cases. 

In general, one observes a rapid relaxation in time, typically described by a 
double exponential form when considering particles on a chain; and by a factorial (rather than
exponential) decrease of the connected spatial correlators. 
In the vicinity of a boundary, the results
deviate strongly from what is found deep in the bulk.  
In comparison with the {\sc kdh} model which originally motivated this work, 
the kink densities and correlators under consideration here 
are distinct from the slowly relaxing global observables examined before 
\cite{Kimball79,Deker79,Dutta09}. It would be interesting to see whether
one might identify a `dual' to the two-species annihilation model 
of which global variables could be treated in a way analogous
to the {\sc kdh}-model. 
The two-species model offers the further possibility to analyse segregation phenomena.  

\vspace{0.5truecm}
\noindent 
{\bf Acknowledgements:}
We thank P.L. Kaprisvky for useful correspondence. 
This collaboration has benefited from an agreement between the 
french COFECUB agency and the university of S\~ao Paulo
(agreement Uc Ph 116/09).

\newpage 
\appsection{A}{Calculational details: single-species model}

\subsection{{Correlator of strings with an one-site hole}}
For spatially translation-invariant initial conditions, the equation of motion of the
averages of the 
single-site hole $\underbrace{\bullet\bullet\ldots\bullet}_{n} \fbox{~1~} 
\underbrace{\bullet\bullet\ldots\bullet}_{m}$ are given in (\ref{24}). 
In the same way as in eq.~(\ref{Trick1}), we set
\BEQ C_n(t)=u_n(s)e^{-2(n-1)t} \;\; , \;\; 
   C_{n,m}^1(t)=u_{n,m}^1(s)e^{-2(n+m-2)t}  \label{Trick1B}
\EEQ
and use again the change of variables (\ref{RenorTime}). The equations of motion (\ref{24}) become 
   \be \label{26} 
   {\D\over \D s}u_{n,m}^1(s)=2\big[u^1_{n+1,m}(s)+u^1_{n,m+1}(s)
      +2(2s+1)u_{n+m+1}(s)\big]
   \ee
and the definition of the $u_{n,m}^1(s)$ are extended such that they hold true for all $n,m\geq 0$. 
We stress that $u_{n,0}^1(s)\ne u_{n}(s) \ne u_{0,n}^1(s)$ 
are unrelated to the density of an unbroken $n$-string, they are rather
a computational device without an obvious physical interpretation. 
Introduce the generating function
   \be F^1(x,y,s):=\sum_{n,m=0}^{\infty} {u_{n,m}^1(s)\over
     n!\ \!m!}x^ny^m\label{DefF1}\ee
and derive its equation of motion as follows
   \ba\fl {\partial\over\partial s} F^1(x,y,s)
   =2\sum_{n,m=0}^{\infty} \big[u^1_{n+1,m}(s)+u^1_{n,m+1}(s)
     +2(2s+1)u_{n+m+1}(s)\big]{x^ny^m\over n!\ \!m!}  \nonumber\\
   \lo=2\sum_{n=1,\atop m=0}^{\infty}  u^1_{n,m}(s){x^{n-1}y^m\over
     (n-1)!m!}+2\sum_{n=0,\atop m=1}^{\infty}  u^1_{n,m}(s){x^ny^{m-1}\over
     n!(m-1)!}\nonumber\\
   \lo\quad\quad+4(2s+1)\sum_{n,m=0}^{\infty} {u_{n+m+1}(s)\over
     (n+m)!}{(n+m)!\over n!m!} x^ny^m\nonumber\\
   \lo=2\left({\partial\over\partial x}+{\partial\over\partial
     y}\right)F^1(x,y,s)+4(2s+1)\sum_{k=0}^{\infty} {u_{k+1}(s)\over
     k!}(x+y)^k\nonumber\\
    \lo=2\left({\partial\over\partial x}+{\partial\over\partial
     y}\right)F^1(x,y,s)+4(2s+1)\sum_{k=1}^{\infty} {u_{k}(s)\over
     (k-1)!}(x+y)^{k-1}
  \ea
The last term involves the derivative of the generating function
$F'(x,s)=\partial_x F(x,s)$, see eq.~(\ref{DefF}), 
of the $n$-string density without a hole. It remains 
   \be \label{29}
   \hspace{-1.0truecm}{\partial\over\partial s} F^1(x,y,s)
   =2\left({\partial\over\partial x}+{\partial\over\partial
     y}\right)F^1(x,y,s)+4(2s+1)F'(x+y,s) \,. \ee
Equations of this kind are conveniently solved by an appropriate change of variables. 
We choose $F^1(s,x,y)=g^1(\alpha,\beta,s)$ where $\alpha=2s+x$ and 
$\beta=2s+y$.\footnote{To find these, recall that the homogeneous
equation $(\partial_s -2\partial_x-2\partial_y)F=0$ admits any function 
$F=f(2s+x,2s+y)$ as solution, see e.g. \cite{Kamke59}.} 
With the chosen variables, eq.~(\ref{29}) becomes an ordinary differential equation 
   \be\fl\left({\partial\over\partial s}-2{\partial\over\partial x}
   -2{\partial\over\partial y}\right)F^1(x,y,s)
   ={\partial g^1\over\partial s}(\alpha,\beta,s)
   =4(2s+1)F'(\alpha+\beta-4s,s)\ee
where $\alpha,\beta$ only enter as parameters. According to eq.~(\ref{eq20}), 
$F'(x,s)=F'(x+4s,0)$ and using eq.~(\ref{FctGeneF1}) for $F$,
it follows
   \BEQ 
   \hspace{-1.5truecm}{\partial g^1\over\partial s}(\alpha,\beta,s) 
   = 4(2s+1) F'(\alpha+\beta,0) = 4(2s+1)\sum_{n=0}^{\infty} {u_{n+1}(0)\over n!}(\alpha+\beta)^n
   \EEQ
Remarkably, the dependence on $s$ is merely the rather simple explicit one, 
such that the integration readily gives 
\BEA  
\hspace{-1.0truecm}       g^1(\alpha,\beta,s)&=& 4(s^2+s)\sum_{n=0}^{\infty}
       {u_{n+1}(0)\over n!}(\alpha+\beta)^n+g^1(\alpha,\beta,0)\nonumber\\
\hspace{-1.0truecm}       &=& 4(s^2+s)\sum_{n=0}^{\infty}{u_{n+1}(0)\over n!}(\alpha+\beta)^n
                              +\sum_{n,m=0}^{\infty} {u_{n,m}^1(0)\over n!\ \!m!}\alpha^n\beta^m
       \label{Solg1}
\EEA
where the integration constant appears as the additional term $g^1(\alpha,\beta,0)$.
Next, express this generating function in terms of the original
variables $x,y$ and $s$:
   \ba \hspace{-2.5truecm}F^1(x,y,s)&=&4(s^2+s)\sum_{n=0}^{\infty}
          {u_{n+1}(0)\over n!}(4s+x+y)^n+\sum_{n,m=0}^{\infty} {u_{n,m}^1(0)\over
         n!\ \!m!}(2s+x)^n(2s+y)^m\nonumber\\
       \hspace{-2.5truecm}&=&4(s^2+s)\sum_{k,l,m=0}^{\infty}{u_{k+l+m+1}(0)\over
         k!\ \!l!\ \!m!}(4s)^kx^ly^m\nonumber\\
       \hspace{-2.5truecm}& & \quad\quad+\sum_{k,l=0}^{\infty}\sum_{n=k}^{\infty}\sum_{m=l}^{\infty}
       {u_{n,m}^1(0)\over k!\ \!(n-k)!\ \!l!\ \!(m-l)!}
       (2s)^{n-k}x^k(2s)^{m-l}y^l
       \ea
The identification with the definition of the generating function
(\ref{DefF1}) leads to
   \be\fl u_{n,m}^1(s)=4(s^2+s)\sum_{k=0}^{\infty}{u_{k+n+m+1}(0)\over
     k!}(4s)^k+\sum_{k,l=0}^{\infty}{u_{n+k,m+l}^1(0)\over k!\ \!l!}
       (2s)^{k+l}\ee
the unique solution of eq.~(\ref{26}). In addition, neither 
$u_{n,0}^{1}(0)$, nor $u_{0,m}^{1}(0)$ nor $u_{0}(0)$ enter into the
observables $u_{n,m}^{1}(s)$ with $n,m\geq 1$, as it should be. 
After performing the changes (\ref{RenorTime}) and (\ref{Trick1}),
the desired correlator becomes 
   \ba C^1_{n,m}(t)&=\left(e^{-4t}-1\right)
   \sum_{k=0}^{\infty} {u_{k+n+m+1}(0)\over k!}2^k\left(e^{-2t}-1\right)^ke^{-2(n+m-2)t}
   \nonumber\\
   &\quad\quad+\sum_{k,l=0}^{\infty}{u_{n+k,m+l}^1(0)\over k!\ \!l!}
       \left(e^{-2t}-1\right)^{k+l}e^{-2(n+m-2)t}
   \ea
Finally, since the first line can be expressed in terms of $C_n(t)$, 
using (\ref{SolFinCorr}), and since $C_{n,m}^{1}(0)=u_{n,m}^1(0)$, the final solution is given
by eq.~(\ref{C:30}) in the main text. 

\subsection{{Correlator of string with a hole of $r$ sites}}
The equation of motion for the $(n,m)$-correlator of two strings of particles (kinks) separated by a
hole consisting of $r>1$ sites, for translation-invariant initial conditions, are, see also \cite{deSm02}
    \be\fl{\D \over \D t}C_{n,m}^r(t)=-2\big[C^r_{n+1,m}(t)+C^r_{n,m+1}(t)
      +(n+m-2)C_{n,m}^r(t)+C_{n+1,m}^{r-1}(t)+C_{n,m+1}^{r-1}(t)\big]
    \ee
Analogously to previous calculations, set 
    \be C_{n,m}^r(t)=u_{n,m}^r(s)e^{-2(n+m-2)t},\quad (r>1)
    \label{ChgCkuk}\ee
and change the time according to (\ref{RenorTime}) such that the 
equations of motion become
    \be \label{41} 
    {\D\over \D s}u_{n,m}^r(s)=2\big[u^r_{n+1,m}(s)+u^r_{n,m+1}(s)
      +u_{n+1,m}^{r-1}(s)+u_{n,m+1}^{r-1}(s)\big]
    \ee
and are extended to all $n,m\geq 0$, which defines the auxiliary quantities $u_{n,0}^r(s)$ and $u_{0,m}^r(s)$. 
Introduce the generating function
    \be F^r(x,y,s):=\sum_{n,m=0}^{\infty} {u_{n,m}^r(s)\over
     n!\ \!m!}x^ny^m\label{DefFk}
    \ee
whose evolution equation is, for $r\geq 2$
    \ba\fl {\partial\over\partial s} F^r(x,y,s)
    =2\sum_{n,m=0}^{\infty} \big[u^r_{n+1,m}(s)+u^r_{n,m+1}(s)
      +u^{r-1}_{n+1,m}(s)+u^{r-1}_{n,m+1}(s)\big]{x^ny^m\over
      n!\ \!m!}\nonumber\\
    \lo=2\sum_{n=1,\atop m=0}^{\infty} u^r_{n,m}(s){x^{n-1}y^m\over
      (n-1)!\ \!m!}+2\sum_{n=0,\atop m=1}^{\infty} u^r_{n,m}(s){x^ny^{m-1}\over
      n!\ \!(m-1)!}\nonumber\\
    \lo\quad+2\sum_{n=1,\atop m=0}^{\infty} u^{r-1}_{n,m}(s){x^{n-1}y^m\over
      (n-1)!\ \!m!}+2\sum_{n=0,\atop m=1}^{\infty} u^{r-1}_{n,m}(s){x^ny^{m-1}\over
      n!\ \!(m-1)!}\nonumber\\
    \lo=2\left({\partial\over\partial x}+{\partial\over\partial
      y}\right)\left(F^r(x,y,s)+F^{r-1}(x,y,s)\right)
    \ea
As before for the case $r=1$, this is solved by changing variables according to 
$\alpha=2s+x$ and $\beta=2s+y$ such that the
function $F^r(x,y,s)=g^r(\alpha,\beta,s)$ obeys the recursion relation 
   \be\fl\left({\partial\over\partial s}-2{\partial\over\partial x}
   -2{\partial\over\partial y}\right)F^r(x,y,s)
   ={\partial g^r\over\partial s}(\alpha,\beta,s)
   =2\left({\partial\over\partial\alpha}+{\partial\over\partial
      \beta}\right)g^{r-1}(\alpha,\beta,s)\ee
We express this recursion in integral form for $g^r(\alpha,\beta,s)$ 
   \be g^r(\alpha,\beta,s)=2\int \!\D s\:\left({\partial\over\partial\alpha}
   +{\partial\over\partial \beta}\right)g^{r-1}(\alpha,\beta,s)
   \label{Integrgk}
   \ee
In order to evaluate this, we consider first a few special cases. 
In the special case $r=2$, the expression (\ref{Solg1}) of
$g^1(\alpha,\beta,s)$ leads to
   \ba g^2(\alpha,\beta,s)
       &=16\left({s^3\over 3}+{s^2\over 2}\right)\sum_{n=0}^{\infty}
       {u_{n+2}(0)\over n!}(\alpha+\beta)^n   \nonumber\\
       &\quad
       +2s\sum_{n,m=0}^{\infty}\left[u_{n+1,m}^1(0)+u_{n,m+1}^1(0)\right]
       {\alpha^n\beta^m\over n!\ \!m!}\nonumber\\
        &\quad +g^2(\alpha,\beta,0)
       \label{Solg2}\ea
since
   \be\fl{\partial\over\partial\alpha}\left[\sum_{n=0}^{\infty}
       {u_{n+1}(0)\over n!}(\alpha+\beta)^n\right]
       =\sum_{n=1}^{\infty} {u_{n+1}(0)\over (n-1)!}(\alpha+\beta)^{n-1}
       =\sum_{n=1}^{\infty} {u_{n+2}(0)\over n!}(\alpha+\beta)^n\ee
Expanding $g^2(\alpha,\beta,0)$ as
     \be g^2(\alpha,\beta,0)=\sum_{n,m=0}^{\infty}u_{n,m}^2(0){\alpha^n\beta^m\over
           n!\ \!m!}\ee
and applying (\ref{Integrgk}) once more, we find 
   \ba\fl g^3(\alpha,\beta,s)
       =64\left(2{s^4\over 4!}+{s^3\over 3!}\right)\sum_{n=0}^{\infty}
       {u_{n+3}(0)\over n!}(\alpha+\beta)^n   \nonumber\\
       \lo\quad
       +2s^2\sum_{n,m=0}^{\infty}\left[u_{n+2,m}^1(0)+2u_{n+1,m+1}^1(0)+u_{n,m+2}^1(0)
         \right]{\alpha^n\beta^m\over n!\ \!m!}\nonumber\\
       \lo\quad
       +2s\sum_{n,m=0}^{\infty}\left[u_{n+1,m}^2(0)+u_{n,m+1}^2(0)\right]
       {\alpha^n\beta^m\over n!\ \!m!}\nonumber\\
        \lo\quad +g^3(\alpha,\beta,0)
       \label{Solg3}\ea
By further iterating the procedure, we conclude that the general result is
   \ba\fl g^r(\alpha,\beta,s)
       =4^r\left(2{s^{r+1}\over (r+1)!}+{s^r\over r!}\right)
       \sum_{n=0}^{\infty}{u_{n+r}(0)\over n!}(\alpha+\beta)^n   \nonumber\\
       \lo\quad +\sum_{l=1}^r {(2s)^{r-l}\over (r-l)!}\sum_{n,m=0}^{\infty}
       \left[\sum_{i=0}^{r-l} {(r-l)!\over i!(r-l-i)!}u_{n+i,m+r-l-i}^l(0)\right]
            {\alpha^n\beta^m\over n!\ \!m!}\ea
and one can indeed check that this solves the recursion (\ref{Integrgk}). 
Next, we express this result in terms of $x$, $y$ and $s$:
\BEA 
   \lefteqn{\hspace{-2.0truecm}F^r(x,y,s)
   =4^r\left(2{s^{r+1}\over (r+1)!}+{s^r\over r!}\right)\sum_{n=0}^{\infty}
   {u_{n+r}(0)\over n!}(4s+x+y)^n}   \nonumber\\
   \hspace{-2.0truecm}& &\hspace{-1.8truecm}+\sum_{l=1}^r (2s)^{r-l}\sum_{n,m=0}^{\infty}
   \left[\sum_{i=0}^{r-l} {u_{n+i,m+r-l-i}^l(0)\over i!\ \!(r-l-i)!}\right]
        {(2s+x)^n(2s+y)^m\over n!\ \!m!} \nonumber\\
   \hspace{-2.0truecm}&\hspace{-2.1truecm}=& 
   \hspace{-1.8truecm}\sum_{a,b=0}^{\infty} \left( 4^r\left(2{s^{r+1}\over (r+1)!}+{s^r\over r!}\right)
   \sum_{c=0}^{\infty}{u_{a+b+c+r}(0)\over \!c!}
   (4s)^c  \right) \frac{x^a y^b}{a!\ b!}   \nonumber\\
   \hspace{-2.0truecm}& & \hspace{-1.8truecm}+ \sum_{a,b=0}^{\infty} \left( \sum_{l=1}^r (2s)^{r-l}
   \sum_{n=a,\atop m=b}^{\infty}\left[\sum_{i=0}^{r-l}
     {u_{n+i,m+r-l-i}^l(0)\over i!\ \!(r-l-i)!}\right]
        {(2s)^{n-a}(2s)^{m-b}\over \!(n-a)!\ \!(m-b)! }\right) \frac{x^a y^b}{a!\ b!}
\EEA
in order to identify the coefficients, after an additional rearrangement of the sums
\BEA
\hspace{-1.0truecm}u_{n,m}^{r}(s) &=& 4^r \left( \frac{s^r}{r!} 
+ 2\frac{s^{r+1}}{(r+1)!}\right) u_{n+m+r}(s) 
\nonumber \\
& & +\sum_{k,l=0}^{\infty} \frac{(2s)^{k+l}}{k!\ l!} \sum_{q=0}^{r-1} \sum_{q'=0}^{q} 
u_{n+k+q-q',m+l+q'}^{r-q}(0) \left(\vekz{q}{q'}\right) 
\frac{(2s)^q}{q!} 
\label{52}
\EEA
For any fixed initial condition, this gives the unique solution of eq.~(\ref{41}). 
The final correlator is obtained after performing the
transformations (\ref{RenorTime}) and then (\ref{ChgCkuk}) 
and relating the initial coefficients $C_{n,m}^{r}(0)=u_{n,m}^{r}(0)$ we recover (\ref{53}) in the main text. 

\subsection{Densities of $n$-strings with inhomogeneous initial conditions}

The evolution equations for the averages $C_{a;n}(t)$ read
    \be {\D \over \D t}C_{a;n}(t)
    =-2(n-1)C_{a;n}(t)-2C_{a-1;n+1}(t)-2C_{a;n+1}(t)
    \ee
They are solved by letting 
    \be C_{a;n}(t)=u_{a;n}(s)\: e^{-2(n-1)t}\ee
and change time according to (\ref{RenorTime}) such that the equations of motion read, 
extended to all $n\geq 0$ which defines $u_{a;0}(s)$
    \be {\D\over \D s}u_{a;n}(s)
    =2u_{a-1;n+1}(s)+2u_{a;n+1}(s)\ee
Assuming for the time being a spatially infinite system, introduce
the generating function
    \be F(p,x,s):=\sum_{n=0}^{\infty}
    \sum_{a=-\infty}^{\infty} {u_{a;n}(s)\over n!} p^a x^n
    \label{DefFxy}\ee
The evolution equation becomes
    \ba\fl {\partial\over\partial s}F(p,x,s)=
    2\sum_{a=-\infty}^{\infty}\sum_{n=0}^{\infty}
    {u_{a-1;n+1}(s)\over n!} p^a x^n
    +2\sum_{a=-\infty}^{\infty}\sum_{n=0}^{\infty}
    {u_{a;,n+1}(s)\over n!} p^a x^n
    \nonumber\\
    \lo=2\sum_{n=0}^{\infty}\sum_{a=-\infty}^{\infty} {u_{a;n}(s)\over n!} p^{a+1}  x^{n-1}
    +2\sum_{n=1}^{\infty}\sum_{a=-\infty}^{\infty} {u_{a;n}(s)\over n!} p^a x^{n-1}
    \nonumber\\
    \lo=2(1+p){\partial\over\partial x}F(p,x,s)
    \ea
where the variable $p$ appears only as a parameter. 
Consequently, and taking the initial condition into account, 
the generating function has the form 
    \be \label{76} F(p,x,s) = F\big(p,2(1+p)s+x,0\big)\ee
Expanding according to (\ref{DefFxy}) and using Newton's multinomial formula
    \ba\fl F(p,x,s) =\sum_{a=-\infty}^{\infty} \sum_{n=0}^{\infty}
    {u_{a;n}(0)\over n!} p^a \big(2(1+p)s+x\big)^n \nonumber\\
    \lo=\sum_{a=-\infty}^{\infty} \sum_{n=0}^{\infty}
    {u_{a;n}(0)\over n!} p^a \Big[\sum_{m,k,l=0}^{\infty} \delta_{n,m+k+l} 
      {n!\over m!\ \!k!\ \!l!}(2s)^m(2ps)^k x^l\Big]     \nonumber\\
    \lo=\sum_{a=-\infty}^{\infty} \sum_{m,k,l=0}^{\infty}
    {u_{a;m+k+l}(0)\over m!\ \!k!\ \!l!} p^a (2s)^{m+k} p^k x^l\nonumber\\
    \lo=\sum_{a=-\infty}^{\infty} \sum_{l=0}^{\infty} \frac{p^a x^l}{l!} \sum_{m,k=0}^{\infty}
    {u_{a-k;m+k+l}(0)\over m!\ \!k!}(2s)^{m+k} 
    \ea
By comparison with eq. (\ref{DefFxy}), we find 
   \be \label{78} 
   u_{a;n}(s)=\sum_{m,k=0}^{\infty}
   {u_{a-k;m+k+n}(0)\over m!\ \!k!}(2s)^{m+k}\ee
and finally obtain (\ref{79}) in the main text. 

The particular case of homogeneous initial conditions (\ref{SolFinCorr}) is recovered when
setting $C_{a-k;m+k+n}(0)=C_{m+k+n}(0)$ and rearranging indices.

\subsection{Correlation of two $n$-strings with inhomogeneous initial conditions}

\subsubsection{\underline{String with an one-site hole}}
Analogously to sections~2 and~A.2, it is useful to consider first a $(n,m)$-correlator of two strings which are
separated by a hole of a single site. Consider the averages $C_{a;n,m}^1(t)$ which 
satisfy the equation of motion
\BEA
\hspace{-1.0truecm}\frac{\D}{\D t} C_{a;n,m}^1(t) &=& - 2(n+m-2) C_{a;n,m}^1(t) \nonumber \\
& & - 2 C_{a-1;n+1,m}^1(t) - 2 C_{a;n,m+1}^1(t) - 4C_{a;n+m+1}(t)
\EEA
natural generalisation of (\ref{24}). Define $C_{a;n,m}^1(t)=u_{a;n,m}^1(s) e^{-2(n+m-2)t}$
and also use (\ref{RenorTime}) to obtain
\BEQ \label{84}
\hspace{-1.9truecm}\frac{\D}{\D s}u_{a;n,m}^1(s) 
= u_{a-1;n+1,m}^1(s) + u_{a;n,m+1}^1(s) +2(2s+1)u_{a;n+m+1}(s)
\EEQ
extended to all $n,m\geq 0$. 
The next step is the definition of the appropriate generating function
\BEQ
\hspace{-1.9truecm}F^1(p,x,y,s) := 
\sum_{a=-\infty}^{\infty} \sum_{n,m=0}^{\infty} \frac{u_{a;n,m}^1(s)}{n!\ m!} p^a x^n y^m
\EEQ
which can be shown, analogously to what has been done above, to satisfy the equation
\BEQ
\hspace{-1.9truecm}\demi \frac{\partial F^1(p,x,y,s)}{\partial s} 
= p \frac{\partial F^1(p,x,y,s)}{\partial x} 
+ \frac{\partial F^1(p,x,y,s)}{\partial y} 
+2(2s+1) F'(p,x+y,s)
\EEQ
where $F=F(p,x,s)$ is defined in (\ref{DefFxy}) and $F'=\partial_x F$. To solve this, we perform the
change of variables
\BEQ
\hspace{-1.9truecm}G^1(p,\alpha,\beta,s) := F^{1}(p,x,y,s)  \;\; , \;\;
\alpha := 2(1+p)s+x+y \;\; , \;\; \beta := x - p y
\EEQ
and find the simplified equation 
\BEQ
\hspace{-1.9truecm}\demi \frac{\partial G^1(p,\alpha,\beta,s)}{\partial s} 
= 2(2s+1) F'(p,\alpha-2(1+p)s,s) = 2(2s+1) F'(p,\alpha,0)
\EEQ
where in the last step eq.~(\ref{76}) was used. Integration gives
\BEQ
G^1(p,\alpha,\beta,s) = 4s(s+1) F'(p,\alpha,0) + G^1(p,\alpha,\beta,0)
\EEQ
and from the initial condition and the definition of $G^1$, we also have
\BEQ
G^1(p,\alpha,\beta,0) = F^1\left( p, \frac{p\alpha+\beta}{1+p}, \frac{\alpha-\beta}{1+p},0\right)
\EEQ
Inserting and converting back to the original variables, 
a straightforward but just a little tedious computation leads to
\BEA
\hspace{-2.0truecm}\lefteqn{F^1(p,x,y,s) = 2s(s+1) \sum_{a=-\infty}^{\infty} 
\sum_{m=0}^{\infty} \frac{u_{a;m+1}(0)}{m!} \left[ 2(1+p)s +x+y\right]^m p^a }
\nonumber \\
\hspace{-2.0truecm}&&+ \sum_{a=-\infty}^{\infty} 
\sum_{n,m=0}^{\infty} \frac{u_{a;n,m}^{1}(0)}{n!\ m!} \left[ 2ps+x\right]^n \left[2s+y\right]^m p^a
\nonumber \\
\hspace{-2.0truecm}&=& \sum_{a=-\infty}^{\infty} \sum_{n,m=0}^{\infty} \frac{p^a x^n y^m}{n!\ m!} \left[ 
\sum_{\ell=0}^{\infty} \frac{(2s)^{\ell}}{\ell!} 
\sum_{k=0}^{\ell} \left(\vekz{\ell}{k}\right) u_{a-k;n+k,m-\ell+k}^{1}(0) \right] 
\\
\hspace{-2.0truecm}&&+2s(s+1) \sum_{a=-\infty}^{\infty} \sum_{n,m=0}^{\infty} \frac{p^a x^n y^m}{n!\ m!} 
\left[ 
\sum_{k=0}^{\infty} \frac{(2s)^{k}}{k!} \sum_{k\ell=0}^{k} \left(\vekz{k}{\ell}\right) u_{a-\ell;n+k+m+1}(0) 
\right] 
\nonumber
\EEA
and from which one reads off the coefficients
\BEA
\lefteqn{u_{a;n,m}^{1}(s) = \sum_{\ell,k=0}^{\infty} \frac{(2s)^{\ell+k}}{\ell!\ k!} 
u_{a-\ell;n+\ell,m+k}^1(0)} \nonumber \\
&&+2s(s+1) \sum_{k=0}^{\infty} \frac{(2s)^{k}}{k!} \sum_{\ell=0}^{k} \left(\vekz{k}{\ell}\right) 
u_{a-\ell;n+k+m+1}(0)
\nonumber \\
&=& \sum_{\ell,k=0}^{\infty} \frac{(2s)^{\ell+k}}{\ell!\ k!} u_{a-\ell;n+\ell,m+k}^1(0) \:+ 2s(s+1) 
u_{a;n+m+1}(s)
\EEA
At this point, it is an useful exercise to check that this is indeed the unique solution 
of the equation of motion (\ref{84}). 

\subsubsection{\underline{String with a hole of $r$ sites}}
The general $(n,m)$-correlator of two strings $C_{a;n,m}^r$, separated by a hole of $r$ sites, 
satisfies the equation of motion, with $r\geq 2$
\BEA
\hspace{-1.0truecm}\lefteqn{\frac{\D}{\D t} C_{a;n,m}^r(t) = - 2(n+m-2) C_{a;n,m}^r(t)} \nonumber \\
& & \hspace{-1.0truecm}- 2 C_{a-1;n+1,m}^r(t) - 2 C_{a;n,m+1}^r(t) 
    -  2 C_{a;n+1,m}^{r-1}(t) - 2 C_{a;n,m+1}^{r-1}(t)
    \label{A:44}
\EEA
This may be simplified as usual by defining $C_{a;n,m}^r(t)=u_{a;n,m}^r(s)\, e^{-2(n+m-2)t}$
and using (\ref{RenorTime}), we obtain, for $r\geq 2$
\BEQ \label{95}
\hspace{-1.9truecm}\frac{\D}{\D s}u_{a;n,m}^r(s) 
= u_{a-1;n+1,m}^r(s) + u_{a;n,m+1}^r(s) + u_{a;n+1,m}^{r-1}(s) + u_{a;n,m+1}^{r-1}(s) 
\EEQ
In principle, we could again define a generating function and solve the corresponding differential equation. 
However, 
the great similarity between the results of the previous sub-section of this appendix  
allows us to write down immediately the expected solution
\BEA
\hspace{-2.0truecm}\lefteqn{u_{a;n,m}^{r}(s) = \sum_{\ell,k=0}^{\infty} \frac{(2s)^{\ell+k}}{\ell!\ k!} 
\sum_{q=0}^{r-1} \sum_{q'=0}^{q} u_{a-\ell;n+\ell+q-q',m+k+q'}^{r-q}(0) \left(\vekz{q}{q'}\right) 
\frac{(2s)^q}{q!} } \nonumber \\
&&+4^r \left( \frac{s^r}{r!} + 2\frac{s^{r+1}}{(r+1)!}\right) u_{a;n+m+r}(s)
\label{A:46}
\EEA
valid for $n,m\in\mathbb{N}$, $a\in\mathbb{Z}$ and all integers $r\geq 1$. Here, we have also used the 
solution (\ref{78}) for 
string-densities without a hole. From this, one readily recovers (\ref{97}) in the main text. 

Indeed, it is a straightforward matter to verify
that the solution (\ref{A:46}) indeed solves eq.~(\ref{95}) and it obviously satisfies the required initial condition 
for $s=0$. 
Hence we have found, for any fixed initial condition, the unique solution of (\ref{95}). 
For a rapid check, note that in the case of spatially translation-invariant initial conditions, 
we recover our previous result (\ref{52}).

\appsection{B}{Calculational details: two-species model}

\subsection{Densities of $n$-strings}

The two string densities $A_n$ and $B_n$ obey the equations (\ref{ab1},\ref{ab2}), which are quite similar to 
(\ref{eqEvolCn}). The method of solution close follows the one used in appendix~A.  
Perform the change of variables 
    \be 
    \hspace{-0.5truecm}
    A_n(t)=u_n(s)e^{-(n-1)t} \;\; , \;\;
    B_n(t)=v_n(s)e^{-(n-1)t} \;\; ; \;\; s := e^{-t}-1 
    \label{Trick1AB}
    \ee
so that the equations of motion become, for all $n\geq 0$ 
    \be
    \hspace{-0.5truecm} 
    {\D\over\D s}u_n(s)=u_{n+1}(s)+v_{n+1}(s) \;\; ,\;\; 
    {\D\over\D s}v_n(s)=v_{n+1}(s)+u_{n+1}(s)
    \ee
These two equations can be decoupled by the further change of variables 
    \be 
    X_n(s):=u_n(s)-v_n(s) \;\; , \;\;  Y_n(s):=u_n(s) +v_n(s)
    \label{DefXYAB}
    \ee
for which the evolution equations read
    \be 
    {\D\over\D s}X_n(s)=0 \;\; , \;\; {\D\over\D s}Y_n(s)=2Y_{n+1}(s)
    \label{EqDecouplXYAB}
    \ee
The first equation gives trivially $X_n(s)=X_n(0)$ while the second one reduces to (\ref{EqEvolun2}), 
up to a time-rescaling so that
we have $Y_n(s)=C_n(s/2)$, already analysed in section~2, with the proper identification of the initial condition. Adapting 
(\ref{SolFinCorr}),  one has
   \ba
   \fl u_n(s)={1\over 2}\big(X_n(s)+Y_n(s)\big)
   ={1\over 2}X_n(0)+{1\over 2}\sum_{m=0}^{\infty} {2^m\over m!}
   Y_{m+n}(0)\left(e^{-t}-1\right)^m\nonumber\\
   \fl v_n(s)={1\over 2}\big(Y_n(s)-X_n(s)\big)
   =-{1\over 2}X_n(0)+{1\over 2}\sum_{m=0}^{\infty} {2^m\over m!}
   Y_{m+n}(0)\left(e^{-t}-1\right)^m
   \label{108}
   \ea
and this finally leads to the result stated in eq.~(\ref{109}). 

As a preparation for the computation of the correlators below, define the generating functions
\BEA
F(x,s) &:=& \sum_{n=0}^{\infty} u_n(s) \frac{x^n}{n!} \;\; , \;\; 
G(x,s) \::=\: \sum_{n=0}^{\infty} v_n(s) \frac{x^n}{n!} \;\; , \;\; \nonumber \\
X(x,s) &:=& \sum_{n=0}^{\infty} X_n(s) \frac{x^n}{n!} \;\; , \;\; 
Y(x,s) \::=\: \sum_{n=0}^{\infty} Y_n(s) \frac{x^n}{n!}  
\label{110}
\EEA
which are to be found by solving the differential equations
\BEQ
\frac{\partial F}{\partial s} = \frac{\partial F}{\partial x} + \frac{\partial G}{\partial x} \:=\: 
\frac{\partial G}{\partial s} 
\;\; , \;\; 
\frac{\partial X}{\partial s} = 0 \;\; , \;\;
\frac{\partial Y}{\partial s} = 2 \frac{\partial Y}{\partial x}
\EEQ
Hence the equations for $X(x,s)$ and $Y(x,s)$ 
decouple and we recognise those already treated in section~2 and appendix~A. We are thus back to (\ref{109}). 

\subsection{Correlations between two strings $ABAB\ldots$ separated by one site}

The equations for two $n$-strings separated by hole of $r=1$ site read 
   \ba\fl&{\D\over\D t}A^1_{n,m}=-B^1_{n+1,m}-(n-1)A^1_{n,m}-2A_{n+m+1}
   -(m-1)A^1_{n,m}-A^1_{n,m+1}\nonumber\\
   \fl&{\D\over\D t}B^1_{n,m}=-A^1_{n+1,m}-(n-1)B^1_{n,m}-2B_{n+m+1}
   -(m-1)B^1_{n,m}-B^1_{n,m+1}\ea
where $A_n$ and $B_n$ are the correlations of a string of particles
without holes previously calculated. Perform the habitual change of variables 
$s=e^{-t}-1$ and 
   \BEQ 
\fl   A^1_{n,m}(t)=u^1_{n,m}(s)\,e^{-(n+m-2)t} \;\; , \;\; 
   B^1_{n,m}(t)=v^1_{n,m}(s)\,e^{-(n+m-2)t}\label{Trick1ABHole}
   \EEQ
while for the non-interrupted $A_n$ and $B_n$, use again (\ref{Trick1AB}). This leads to 
   \ba
   {\D\over\D s}u^1_{n,m}(s)=v^1_{n+1,m}(s)+2(s+1)u_{n+m+1}(s)+u^1_{n,m+1}(s)\nonumber\\
     {\D\over\D s}v^1_{n,m}(s)=u^1_{n+1,m}(s)+2(s+1)v_{n+m+1}(s)+v^1_{n,m+1}(s)
   \ea

This set of equations cannot be decoupled at this point by defining
linear combinations of them as in (\ref{DefXYAB}).
We therefore introduce the generating functions
   \be\fl 
   F^1(x,y,s):=\sum_{n,m=0}^{\infty} u^1_{n,m}(s){x^ny^m\over n!m!} \;\; , \;\; 
     G^1(x,y,s):=\sum_{n,m=0}^{\infty} v^1_{n,m}(s){x^ny^m\over n!m!}
     \ee
and recall the definition (\ref{110}) of $F(x,s)$ and $G(x,s)$ for the correlation without hole.
Multiplying the evolution equations by ${x^ny^m\over n!m!}$
and using the fact that
  \be\fl\sum_{n,m=0}^{\infty} u_{n+m+1}{x^ny^m\over n!m!}
  =\sum_{k=0}^{\infty} u_{k+1}{(x+y)^k\over k!}
  =\sum_{k=1}^{\infty} u_k{(x+y)^{k-1}\over (k-1)!}
  =\left[\sum_{k=0}^{\infty} u_k{(x+y)^k\over k!}\right]'\ee
we obtain (with $F'(x,s)=\partial_x F(x,s)$ and $G'(x,s) = \partial_x G(x,s)$)
   \ba{\partial F^1\over\partial s}={\partial G^1\over\partial x}
   +2(s+1)F'(x+y)+{\partial F^1\over\partial y}\nonumber\\
   {\partial G^1\over\partial s}={\partial F^1\over\partial x}
   +2(s+1)G'(x+y)+{\partial G^1\over\partial y}\ea
Letting now
   \ba X^1(x,y,s)=F^1(x,y,s)-G^1(x,y,s),\nonumber\\
   Y^1(x,y,s)=F^1(x,y,s)+G^1(x,y,s)\ea
the set of equations decouples
   \ba \partial_sX^1=(-\partial_x+\partial_y)X^1+2(s+1)X'(x+y,s)
   \nonumber\\
   \partial_sY^1=(\partial_x+\partial_y)Y^1+2(s+1)Y'(x+y,s)
   \label{EqDiffCorrXYAB}\ea
where $X(x,s)=F(x,s)-G(x,s)$ and $Y(x,s)=F(x,s)+G(x,s)$. The solutions
of the homogeneous PDEs are
    \BEQ
    X^1(x,y,s)=X^1(s-x,s+y) \;\; , \;\; 
    Y^1(x,y,s)=Y^1(s+x,s+y)
    \EEQ
In the following, denote $\alpha_\pm:=s\pm x$ and $\beta:=s+y$ and
perform the following change of variables 
    \be 
    X^1(x,y,s)=x^1(\alpha_-,\beta,s) \;\; , \;\; 
    Y^1(x,y,s)=y^1(\alpha_+,\beta,s)
    \ee
Since
   \be\fl\partial_s X^1-(-\partial_x+\partial_y)X^1
   =\partial_s x^1+\left({\D \alpha_-\over\D s}
   +{\D \alpha_-\over\D x}\right)\partial_{\alpha_-}x^1
   +\left({\D \beta\over\D s}-{\D\beta\over\D y}\right)
   \partial_{\beta}x^1=\partial_s x^1
   \label{partial_sX1}\ee
and
   \ba X'(x+y,s)
   &={\D\over\D x}\sum_{n=0}^{\infty} X_n(0){(x+y)^n\over n!}\nonumber\\
   &=\sum_{n=1}^{\infty} X_n(0){(x+y)^{n-1}\over (n-1)!}\nonumber\\
   &=\sum_{n=0}^{\infty} X_{n+1}(0){(x+y)^n\over n!}\nonumber\\
   &=\sum_{n=0}^{\infty} X_{n+1}(0){(-\alpha_-+\beta)^n\over n!}\nonumber\\
   &=\sum_{n,m=0}^{\infty} (-1)^nX_{n+m+1}(0){\alpha_-^n\beta^m\over
   n!\ \!m!}\ea
because $X_n$ does not depend on $s$ according to
(\ref{EqDecouplXYAB}), the equation of motion for $x^1$ becomes
  \be 
  \partial_s x^1(\alpha_-,\beta,s)=2(s+1)\sum_{n,m=0}^{\infty}
  (-1)^nX_{n+m+1}(0){\alpha_-^n\beta^m\over n!\ \!m!}\,.
  \ee
This is readily integrated and gives
  \be \fl
  x^1(\alpha_-,\beta,s)=(s^2+2s)\sum_{n,m=0}^{\infty}
  (-1)^nX_{n+m+1}(0){\alpha_-^n\beta^m\over n!\ \!m!}
  +x^1(\alpha_-,\beta,0)  \label{SolX1AB}
  \ee
The last term is written as a series
    \be x^1(\alpha_-,\beta,0)=\sum_{n,m=0}^{\infty}
    (-1)^nX^1_{n,m}(0){\alpha_-^n\beta^m\over n!\ \!m!}\ee
The presence of the oscillating factor $(-1)^n$ is required for
$X^1_{n,m}(0)$ to be equal to the initial value of $X^1$, as can be
verified {\sl a-posteriori}. Its origin can be found in the fact that
$\alpha_-=s-x$ leading to a factor $(-x)^n$ in the series when $t$
and thus $s$ vanishes. The generating function $X^1$ is now
   \ba\fl X^1(x,y,s)=x^1(s-x,s+y,s)  \\
    \fl\quad=\sum_{n,m=0}^{\infty}(-1)^n\left[(s^2+2s)X_{n+m+1}(0)
     +X^1_{n,m}(0)\right]{(s-x)^n(s+y)^m\over n!\ \!m!}  \nonumber\\
    \fl\quad=\sum_{n,m,k,l=0}^{\infty}(-1)^{n+k}\left[(s^2+2s)X_{n+m+k+l+1}(0)
     +X^1_{n+k,m+l}(0)\right]{s^{k+l}(-x)^ny^m\over
     n!\ \!k!\ \!m!\ \!l!}\nonumber\\
    \fl\quad=\sum_{n,m,k,l=0}^{\infty}\left[(s^2+2s)X_{n+m+k+l+1}(0)
      {(-s)^ks^l\over k!\ \!l!}+(-1)^kX^1_{n+k,m+l}(0){s^{k+l}\over k!\ \!l!}
      \right]{x^ny^m\over n!\ \!m!\ \!}\nonumber\\
    \fl\quad=\sum_{n,m=0}^{\infty}\left[(s^2+2s)
      X_{n+m+1}(0)+\sum_{k,l=0}^{\infty} (-1)^kX^1_{n+k,m+l}(0)
      {s^{k+l}\over k!\ \!l!}\right]{x^ny^m\over
      n!\ \!m!\ \!}\nonumber
    \label{SolXTrouAB}
   \ea
where we have used the fact that
   \be\fl \sum_{k,l=0}^{\infty} a_{k+l} {(-s)^ks^l\over k!\ \!l!}
   =\sum_{n=0}^{\infty} {a_n\over n!} \sum_{k=0}^n {n!\over
     k!\ \!(n-k)!}(-s)^ks^{n-k}=\sum_{n=0}^{\infty} {a_n\over n!}
    0^n=a_0\label{BinomToZero}\ee

Similarly, the second equation (\ref{EqDiffCorrXYAB}) gives
   \ba\partial_s y^1&=2(s+1)\sum_{n=0}^{\infty}
   Y_{n+1}(s){(x+y)^n\over n!}\nonumber\\
   &=2(s+1)\sum_{n,l=0}^{\infty} Y_{n+l+1}(0)
   {(2s)^l(x+y)^n\over n!\ \!l!}\nonumber\\
   &=2(s+1)\sum_{n,l=0}^{\infty} Y_{n+l+1}(0)
   {(2s)^l(\alpha_++\beta-2s)^n\over n!\ \!l!}\nonumber\\
   &=2(s+1)\sum_{n,m,k,l=0}^{\infty} Y_{n+m+k+l+1}(0)
   {(2s)^l\alpha_+^n\beta^m(-2s)^k\over n!\ \!m!\ \!k!\ \!l!}\nonumber\\
   &=2(s+1)\sum_{n,m=0}^{\infty} Y_{n+m+1}(0)
   {\alpha_+^n\beta^m\over n!\ \!m!}
   \ea
where we have used (\ref{108}). Again the $s$-dependence
disappears and the integration leaves
   \be y^1(\alpha_+,\beta,s)=(s^2+2s)\sum_{n,m=0}^{\infty}
   Y_{n+m+1}(0){\alpha_+^n\beta^m\over n!\ \!m!}
   +y^1(\alpha_+,\beta,0)\label{SolY1AB}\ee
The last term is written as a series
    \be y^1(\alpha_+,\beta,0)=\sum_{n,m=0}^{\infty}
    Y^1_{n,m}(0){\alpha_+^n\beta^m\over n!\ \!m!}
    \ee
The generating function $Y^1$ is thus
   \ba\fl Y^1(x,y,s)=y^1(s+x,s+y,s)  \\
   \fl\quad=\sum_{n,m=0}^{\infty}\left[(s^2+2s)
     Y_{n+m+1}(0)+Y^1_{n,m}(0)\right]
     {(s+x)^n(s+y)^m\over n!\ \!m!}\nonumber\\
    \fl\quad=\sum_{n,m,k,l=0}^{\infty}\left[(s^2+2s)
     Y_{n+m+k+l+1}(0)+Y^1_{n+k,m+l}(0)\right]
     {s^{k+l}x^ny^m\over n!\ \!m!\ \!k!\ \!l!}\nonumber\\
     \fl\quad=\sum_{n,m=0}^{\infty}
     \left[(s^2+2s)\sum_{k=0}^{\infty} Y_{n+m+k+1}(0){(2s)^k\over
         k!}+\sum_{k,l=0}^{\infty}Y^1_{n+k,m+l}(0){s^{k+l}\over
        k!\ \!l!}\right]{x^ny^m\over  n!\ \!m!}\nonumber
     \label{SolYTrouAB}\ea
Taking the sum and the difference of (\ref{SolXTrouAB})
and (\ref{SolYTrouAB}), we recover the generating functions $F^1$ and
$G^1$ and then the coefficients $u^1_{k,l}$ and $v^1_{k,l}$ by identifying
the coefficients of ${x^ny^m\over n!\ \!m!}$:
   \ba\fl
   u^1_{n,m}(s)={1\over 2}\big(X_{n,m}(s)+Y_{n,m}(s)\big)
   = \frac{s^2+2s}{2}\left[X_{n+m+1}(0)+\sum_{k=0}^{\infty}
     Y_{n+m+k+1}(0){(2s)^k\over k!}\right]\nonumber\\
   \lo\quad\quad\quad\quad\quad\quad\quad
   +{1\over 2}\sum_{k,l=0}^{\infty}\left((-1)^kX^1_{n+k,m+l}(0)
   +Y^1_{n+k,m+l}(0)\right){s^{k+l}\over k!\ \!l!}\nonumber\\\ea
and
   \ba\fl
   v^1_{n,m}(s)={1\over 2}\big(Y_{n,m}(s)-X_{n,m}(s)\big)
   =\frac{s^2+2s}{2}\left[-X_{n+m+1}(0)+\sum_{k=0}^{\infty}
     Y_{n+m+k+1}(0){(2s)^k\over k!}\right]\nonumber\\
   \lo\quad\quad\quad\quad\quad\quad\quad
   +{1\over 2}\sum_{k,l=0}^{\infty}\left((-1)^{k+1}X^1_{n+k,m+l}(0)
   +Y^1_{n+k,m+l}(0)\right){s^{k+l}\over k!\ \!l!}\nonumber\\\ea
Applying now the transformations $s=e^{-t}-1$ and
(\ref{Trick1ABHole}), we finally obtain
   \ba\fl
   A^1_{n,m}(t)={1\over 2}\left[\big(e^{-t}-1\big)\big(e^{-t}+1\big)
     \left(X_{n+m+1}(0)+\sum_{k=0}^{\infty}Y_{n+m+k+1}(0)
          {2^k\big(e^{-t}-1\big)^k\over k!}\right)\right.\nonumber\\
   \fl\quad\quad +\left.\sum_{k,l=0}^{\infty}\left((-1)^kX^1_{n+k,m+l}(0)
   +Y^1_{n+k,m+l}(0)\right){\big(e^{-t}-1\big)^{k+l}\over k!\ \!l!}
   \right]e^{-(n+m-2)t}\ea
and
   \ba\fl
   B^1_{n,m}(t)={1\over 2}\left[\big(e^{-t}-1\big)\big(e^{-t}+1\big)
     \left(-X_{n+m+1}(0)+\sum_{k=0}^{\infty}Y_{n+m+k+1}(0)
          {2^k\big(e^{-t}-1\big)^k\over k!}\right)\right.\nonumber\\
   \fl\quad\quad+\left.\sum_{k,l=0}^{\infty}\left((-1)^{k+1}X^1_{n+k,m+l}(0)
   +Y^1_{n+k,m+l}(0)\right){\big(e^{-t}-1\big)^{k+l}\over k!\ \!l!}
   \right]e^{-(n+m-2)t}\ea
as stated in the text. 

\subsection{Correlations between two strings $ABAB\ldots$ separated by $r$ sites}

The equations of motion for the probability to observe two strings separated by a
hole of $r\geq 2$ sites are 
   \ba\fl
   &{\D\over\D t}A^r_{n,m}=-B^r_{n+1,m}-(n-1)A^r_{n,m}-A^{r-1}_{n+1,m}
     -A^{r-1}_{n,m+1}-(m-1)A^r_{n,m}-A^r_{n,m+1}\nonumber\\
   \fl &{\D\over\D t}B^r_{n,m}=-A^r_{n+1,m}-(n-1)B^r_{n,m}-B^{r-1}_{n+1,m}
   -B^{r-1}_{n,m+1}-(m-1)B^r_{n,m}-B^r_{n,m+1}\nonumber \\
   \fl & ~
   \ea
A first simplification is obtained by making the change of variables 
   \BEQ \fl
   A^r_{n,m}(t)=u^r_{n,m}(s)e^{-(n+m-2)t}\;\; , \;\;
   B^r_{n,m}(t)=v^r_{n,m}(s)e^{-(n+m-2)t} \;\; ; \;\; s = e^{-t}-1 \label{TrickkABHole}
   \EEQ
and we have 
   \ba
   {\D\over\D s}u^r_{n,m}(s)=v^r_{n+1,m}(s)+u^{r-1}_{n+1,m}(s)
     +u^{r-1}_{n,m+1}(s)+u^r_{n,m+1}(s)\nonumber\\
     {\D\over\D s}v^r_{n,m}(s)=u^r_{n+1,m}(s)+v^{r-1}_{n+1,m}(s)
        +v^{r-1}_{n,m+1}(s)+v^r_{n,m+1}(s)
   \ea
As before, we assume these equations to be valid for all $n,m\geq 0$ and emphasise that 
$u_{n,0}^r \ne u_{n}^r \ne u_{0,n}^r$ and 
similarly for the $v$. 
We then can introduce the generating functions
   \be\fl 
   F^r(x,y,s)=\sum_{n,m=0}^{\infty} u^r_{n,m}(s)\,{x^ny^m\over n!m!} \;\;,\;\;
     G^r(x,y,s)=\sum_{n,m=0}^{\infty} v^r_{n,m}(s)\,{x^ny^m\over n!m!}
   \ee
Multiplying the evolution equations by ${x^ny^m\over n!m!}$, we obtain
   \ba 
   \partial_s F^r=\partial_x G^r+\partial_y F^r
   +(\partial_x+\partial_y)F^{r-1}\nonumber\\
   \partial_s G^r=\partial_x F^r+\partial_y G^r
   +(\partial_x+\partial_y)G^{r-1}
   \ea
Letting now
   \ba X^r(x,y,s):=F^r(x,y,s)-G^r(x,y,s),\nonumber\\
   Y^r(x,y,s):=F^r(x,y,s)+G^r(x,y,s)\ea
the set of equations decouples
   \ba \partial_sX^r=(-\partial_x+\partial_y)X^r
   +(\partial_x+\partial_y)X^{r-1}\nonumber\\
   \partial_sY^r=(\partial_x+\partial_y)Y^r
   +(\partial_x+\partial_y)Y^{r-1}
   \label{EqDiffkCorrXYAB}\ea
As before in the case $r=1$, and following standard techniques \cite{Kamke59}, 
we denote $\alpha_\pm:=s\pm x$ and $\beta:=s+y$
and we perform the following change of variables 
    \be X^r(x,y,s)=x^r(\alpha_-,\beta,s) \;\; , \;\;
    Y^r(x,y,s)=y^r(\alpha_+,\beta,s)
    \ee

Since the relation (\ref{partial_sX1}) holds also for $X^r$, the PDE
for $x^r$ is
    \be\partial_s x^r(\alpha_-,\beta,s)
    =(\partial_x+\partial_y)X^{r-1}
    =(-\partial_{\alpha_-}+\partial_\beta)x^{r-1}
    \ee
The case $r=2$ is obtained from (\ref{SolX1AB}):
   \ba\fl\partial_sx^2(\alpha_-,\beta,s)\nonumber\\
   \fl=(-\partial_{\alpha_-}+\partial_\beta)\sum_{n,m=0}^{\infty}\left[(s^2+2s)
     (-1)^nX_{n+m+1}(0)+(-1)^nX^1_{n,m}(0)\right]{\alpha_-^n\beta^m\over n!\ \!m!}
   \\
   \fl=\sum_{n,m=0}^{\infty}\left[2(s^2+2s)(-1)^nX_{n+m+2}(0)
     -(-1)^{n+1}X^1_{n+1,m}(0)+(-1)^nX^1_{n,m+1}(0)\right]
     {\alpha_-^n\beta^m\over n!\ \!m!}\nonumber
   \ea
which leads after integration to
   \ba\fl x^2(\alpha_-,\beta,s)=x^2(\alpha_-,\beta,0)\nonumber\\
   \fl+\sum_{n,m=0}^{\infty}(-1)^n\left[2\left({s^3\over 3}+s^2\right)
     X_{n+m+2}(0)+X^1_{n+1,m}(0)s+X^1_{n,m+1}(0)s\right]
           {\alpha_-^n\beta^m\over n!\ \!m!}\ea
where
   \be x^2(\alpha_-,\beta,0)=\sum_{n,m=0}^{\infty} (-1)^nX^2_{n,m}(0)
       {\alpha_-^n\beta^m\over n!\ \!m!}\ee
We can infer the solution for a general $r$~:
   \ba\fl 
   x^r(\alpha_-,\beta,s)
   =\sum_{n,m=0}^{\infty} 2^r\left({s^{r+1}\over (r+1)!}+{s^r\over
     r!}\right)(-1)^nX_{n+m+r}(0){\alpha_-^n\beta^m\over n!\ \!m!}\nonumber\\
   \lo+\sum_{n,m=0}^{\infty}(-1)^n\sum_{l=0}^{k-1}\sum_{p=0}^l
   {l!\over p!\ \!(l-p)!}X^{r-l}_{n+p,m+l-p}(0)
           {s^l\over l!}{\alpha_-^n\beta^m\over n!\ \!m!}
   \ea
Expressing now this quantity in terms of the original variable $x$ and
$y$ leads to
   \ba\fl X^r(x,y,s)=x^r(s-x,s+y,s)\nonumber\\
   \lo=\sum_{n,m=0}^{\infty} 2^r\left({s^{r+1}\over (r+1)!}+{s^r\over
     r!}\right)(-1)^nX_{n+m+r}(0){(s-x)^n(s+y)^m\over n!\ \!m!}\nonumber\\
   \lo+\sum_{n,m=0}^{\infty} (-1)^n\sum_{l=0}^{r-1}\sum_{p=0}^l
           {l!\over p!\ \!(l-p)!}X^{r-l}_{n+p,m+l-p}(0)
           {s^l\over l!}{(s-x)^n(s+y)^m\over n!\ \!m!}\nonumber\\
   \lo=\sum_{n,m,\atop a,b=0}^{\infty} 2^r\left({s^{r+1}\over (r+1)!}+{s^r\over
     r!}\right)(-1)^aX_{n+m+a+b+r}(0){s^{a+b}x^ny^m\over n!\ \!m!\ \!a\ \!b!}
   \nonumber\\
   \lo+\sum_{n,m,\atop a,b=0}^{\infty} (-1)^a\sum_{l=0}^{r-1}\sum_{p=0}^l
    {l!\over p!\ \!(l-p)!}X^{r-l}_{n+p+a,m+l-p+b}(0)
           {s^l\over l!}{s^{a+b}x^ny^m\over n!\ \!a!\ \!m!\ \!b!}\nonumber\\
   \lo=\sum_{n,m=0}^{\infty} 2^r\left({s^{r+1}\over (r+1)!}+{s^r\over
     r!}\right)X_{n+m+r}(0){x^ny^m\over n!\ \!m!}\nonumber\\
   \lo+\sum_{n,m,\atop a,b=0}^{\infty}(-1)^a\sum_{l=0}^{r-1}\sum_{p=0}^l
    {l!\over p!\ \!(l-p)!}X^{r-l}_{n+p+a,m+l-p+b}(0)
           {s^l\over l!}{s^{a+b}x^ny^m\over n!\ \!a!\ \!m!\ \!b!}
   \ea
where we have used (\ref{BinomToZero}).

The same procedure has now to be applied to the calculation of $y^r$.
Similarly to (\ref{partial_sX1}), the PDE for $y^r$ is
    \be
    \partial_sy^r(\alpha_+,\beta,s)
    =(\partial_x+\partial_y)Y^{r-1}
    =(\partial_{\alpha_+}+\partial_\beta)y^{r-1}
    \ee
The case $r=2$ is obtained from (\ref{SolY1AB}):
   \ba\fl
   \partial_sy^2(\alpha_+,\beta,s)
   =(\partial_{\alpha_+}+\partial_\beta)\sum_{n,m=0}^{\infty}
   \left[(s^2+2s)Y_{n+m+1}(0)+Y^1_{n,m}(0)\right]
        {\alpha_+^n\beta^m\over n!\ \!m!}\nonumber\\
   =\sum_{n,m=0}^{\infty}\left[2(s^2+2s)Y_{n+m+2}(0)
     +Y^1_{n+1,m}(0)+Y^1_{n,m+1}(0)\right]{\alpha_+^n\beta^m\over n!\ \!m!}
   \ea
which leads after integration to
   \ba\fl y^2(\alpha_+,\beta,s)=y^2(\alpha_+,\beta,0)\nonumber\\
   \fl\quad\quad+\sum_{n,m=0}^{\infty}\left[2\left({s^3\over 3}+s^2\right)
     Y_{n+m+2}(0)+Y^1_{n+1,m}(0)s+Y^1_{n,m+1}(0)s\right]
           {\alpha_+^n\beta^m\over n!\ \!m!}\ea
where
   \be y^2(\alpha_+,\beta,0)=\sum_{n,m=0}^{\infty} Y^2_{n,m}(0)
       {\alpha_+^n\beta^m\over n!\ \!m!}\ee
We can infer the solution for a general $r$~:
   \ba\fl Y^r(\alpha_+,\beta,s)
   =\sum_{n,m=0}^{\infty} 2^r\left({s^{r+1}\over (r+1)!}+{s^r\over
     r!}\right)Y_{n+m+r}(0){\alpha_+^n\beta^m\over n!\ \!m!}\nonumber\\
   \lo+\sum_{n,m=0}^{\infty}\sum_{l=0}^{r-1}\sum_{p=0}^l {l!\over
     p!\ \!(l-p)!}Y^{r-l}_{n+p,m+l-p}(0){s^l\over l!}
           {\alpha_+^n\beta^m\over n!\ \!m!}\ea
Expressing now this quantity in terms of the original variables $x$ and
$y$ leads to
   \ba\fl Y^r(x,y,s)=y^r(s+x,s+y,s)\nonumber\\
   \lo=\sum_{n,m=0}^{\infty} 2^r\left({s^{r+1}\over (r+1)!}+{s^k\over
     r!}\right)Y_{n+m+r}(0){(s+x)^n(s+y)^m\over n!\ \!m!}\nonumber\\
   \lo+\sum_{n,m=0}^{\infty}\sum_{l=0}^{r-1}\sum_{p=0}^l {l!\over
     p!\ \!(l-p)!}Y^{k-l}_{n+p,m+l-p}(0){s^l\over
     l!}{(s-x)^n(s+y)^m\over n!\ \!m!}
   \nonumber\\
   \lo=\sum_{n,m,c=0}^{\infty} 2^r\left({s^{r+1}\over (r+1)!}+{s^r\over
     r!}\right)Y_{n+m+c+r}(0){(2s)^cx^ny^m\over n!\ \!m!\ \!c!}\nonumber\\
   \lo+\sum_{n,m,\atop a,b=0}^{\infty}\sum_{l=0}^{r-1}\sum_{p=0}^l
           {l!\over p!\ \!(l-p)!} Y^{r-l}_{n+p+a,m+l-p+b}(0)
           {s^l\over l!}{s^{a+b}x^ny^m\over n!\ \!a!\ \!m!\ \!b!}\ea
  
Finally, the original generating functions $F^r={1\over 2}(X^r+Y^r)$
and $G^r={1\over 2}(-X^r+Y^r)$ can be formed and the coefficients
$u_{n,m}^r$ and $v_{n,m}^r$ extracted from their series expansions.
We find
\newpage
\typeout{ *** ici saut de page *** }
   \ba\fl u^r_{n,m}={1\over 2}\left[
     2^r\left({s^{r+1}\over (r+1)!}+{s^r\over r!}\right)
     \Big[X_{n+m+r}(0)+\sum_{c=0}^{\infty} Y_{n+m+c+r}(0){(2s)^c\over c!}\Big]\right.\\
     \fl+\left.\sum_{a,b=0}^{\infty}\sum_{l=0}^{r-1}\sum_{p=0}^l
              {l!\over p!\ \!(l-p)!}\big[(-1)^{a}X^{r-l}_{n+p+a,m+l-p+b}(0)
                +Y^{r-l}_{n+p+a,m+l-p+b}(0)\big]
              {s^{l+a+b}\over l!\ \!a!\ \!b!}\right]\nonumber\\
   \fl v^r_{n,m}={1\over 2}\left[
     2^r\left({s^{r+1}\over (r+1)!}+{s^r\over r!}\right)
     \Big[-X_{n+m+r}(0)+\sum_{c=0}^{\infty} Y_{n+m+c+r}(0){(2s)^c\over c!}
       \Big]\right.\nonumber\\
     \fl+\left.\sum_{a,b=0}^{\infty}\sum_{l=0}^{r-1}\sum_{p=0}^l
              {l!\over p!\ \!(l-p)!}\big[(-1)^{a+1}X^{r-l}_{n+p+a,m+l-p+b}(0)
                +Y^{r-l}_{n+p+a,m+l-p+b}(0)\big]
              {s^{l+a+b}\over l!\ \!a!\ \!b!}\right]\nonumber
\ea
Re-expressing this in terms of the desired correlators, 
we arrive at eqs.~(\ref{166},\ref{167}) in the main text. 

\appsection{C}{Calculational details: single-species model on the Bethe lattice}

The $n$-string averages 
$C_n(t):=\langle \eta_i(t)\eta_j(t) \eta_k(t)
\ldots\rangle$ satisfy the equations of motion
    \be {\D\over \D t}C_n(t)=-2(n-1)C_n-2\left(n(z-2)+2\right) C_{n+1}\ee
To solve these, let
    \be C_n(t)=u_n(s)e^{-2(n-1)t}   \label{Trick1Bethe}\ee
and with the change of variables (\ref{RenorTime}) we have the equations of motion, for all $n\geq 0$
    \be {\D\over \D s}u_n(s)=2\left(n(z-2)+2\right) u_{n+1}(s)
    \label{EqEvolun2Bethe}\ee
As before, introduce the generating function
   \be F(x,s) :=\sum_{n=0}^{\infty} {u_n(s)\over n!}x^n
   \label{DefFBethe}
   \ee
Applying the equation of motion (\ref{EqEvolun2Bethe}) gives the equation 
    \ba\fl {\partial\over \partial s}F(x,s)
    =2\sum_{n=0}^{\infty} [n(z-2)+2]{u_{n+1}(s)\over n!}x^n\nonumber\\
    \lo=2\sum_{n=1}^{\infty} [(n-1)(z-2)+2]{u_n(s)\over
      (n-1)!}x^{n-1}\nonumber\\
    \lo=2\left[(z-2)\sum_{n=1}^{\infty}u_n(s){x^{n-1}\over(n-2)!}
      +2\sum_{n=1}^{\infty}u_n(s){x^{n-1}\over(n-1)!}\right]\nonumber\\
    \lo=2\left[(z-2)x{\partial^2\over \partial x^2}
      +2{\partial\over \partial x}\right]F(x,s)
      \label{61}
    \ea
This PDE is solved by separation ansatz $F(x,s)=f(x)g(s)$
so that eq.~(\ref{61}) becomes
    \be{g'(s)\over g(s)}=2\left[(z-2)x{f''(x)\over f(x)}
      +2{f'(x)\over f(x)}\right]=-\lambda\ee
leading to $g(s)=e^{-\lambda s}$ and an equation for $f(x)$
    \be 2(z-2)xf''(x)+4f'(x)+\lambda f(x)=0\ee
that can be recast as a Bessel equation whose solution is
    \be\fl f(x)=B_\lambda x^{{1\over 2}\left({z-4\over z-2}\right)}
    J_{-{z-4\over z-2}}\left(\sqrt{2\lambda x\over z-2}\right)
    +C_\lambda x^{{1\over 2}\left({z-4\over z-2}\right)}
    Y_{-{z-4\over z-2}}\left(\sqrt{2\lambda x\over z-2}\right)\ee
where $J$ and $Y$ are the Bessel functions of the first and second
kinds. The Bessel function $z^{-\nu}Y_{\nu}(z)$ is always non-analytic at the origin $z=0$. 
Such a behaviour is incompatible with the analyticity built 
into the generating function $F(x,s)$ by construction. 
Hence we must have $C_\lambda=0$ for any $\lambda$. Moreover, negative values of
$\lambda$ are not allowed since they would induce a complex
generating function. It remains
   \be F(x,s)=\int_0^{\infty} \!\D\lambda\: B_\lambda x^{{1\over 2}\left(1-{2\over z-2}\right)}
    J_{-{z-4\over z-2}}\left(\sqrt{2\lambda x\over
      z-2}\right)e^{-\lambda s}
   \ee
Introducing the series expansion of the Bessel function
   \ba\fl F(x,s)=\int_0^{\infty} \!\D\lambda\: B_\lambda
   x^{{z-4\over 2(z-2)}}\left[{\lambda x
      \over 2(z-2)}\right]^{{4-z\over 2(z-2)}}
    \left[\sum_{n=0}^{\infty} {(-1)^n\over n!\Gamma\left(n+{2\over z-2}\right)}
        \left({\lambda x\over 2(z-2)}\right)^n\right]e^{-\lambda s}
    \nonumber\\
    \lo = \sum_{n=0}^{\infty} \frac{x^n}{n!}  \int_0^{\infty} \!\D\lambda\: B_\lambda 
    {(-1)^n\over \Gamma\left(n+{2\over z-2}\right)}
    \left({\lambda\over 2(z-2)}\right)^{n+{1\over z-2}-{1\over 2}}
    e^{-\lambda s}  
    \ea
we can identify the coefficients $u_n$ of the generating function
   \be\fl u_n(s)={(-1)^n\over \Gamma\left(n+{2\over z-2}\right)
   [2(z-2)]^{n+{1\over z-2}-{1\over 2}}}\int_0^{\infty} 
   \!\D\lambda\: B_\lambda\lambda^{n+{1\over z-2}-{1\over 2}}
   e^{-\lambda s}   \label{CoeffBethe1}
   \ee
We note that, since $s\in[-1;0]$, the convergence of the integral requires
that $B_\lambda$ decays sufficiently fast (at least exponentially). In the following,
we shall relate the $B_\lambda$ to the initial conditions $u_n(0)$.
First, note that the initial time coefficient $u_n(0)$ is given by
   \be \fl u_n(0)={(-1)^n\over\Gamma\left(n+{2\over z-2}\right)
   [2(z-2)]^{n+{1\over z-2}-{1\over 2}}}\int_0^{\infty} \!\D\lambda \: B_\lambda\:\lambda^{n+{1\over z-2}-{1\over 2}}
     \label{CoeffBethe2}
   \ee
Expanding now the exponential in the time-dependent coefficients (\ref{CoeffBethe1})
and identifying $u_n(0)$, it follows
   \ba
   \fl u_n(s)={(-1)^n\over \Gamma\left(n+{2\over z-2}\right)
   [2(z-2)]^{n+{1\over z-2}-{1\over 2}}}\int_0^{\infty} \!\D\lambda \: B_\lambda\:\lambda^{n+{1\over z-2}-{1\over 2}}
   \left[\sum_{k=0}^{\infty} {(-\lambda s)^k\over k!}\right]\nonumber\\
   \lo=\sum_{k=0}^{\infty}{(-1)^{n+k}s^k\over k!\Gamma\left(n+{2\over z-2}\right)
   [2(z-2)]^{n+{1\over z-2}-{1\over 2}}}\int_0^{\infty} \!\D\lambda \: B_\lambda\:\lambda^{n+k+{1\over z-2}-{1\over 2}}
   \nonumber\\
   \lo=\sum_{k=0}^{\infty} {[2(z-2)]^k\over k!}{\Gamma\left(n+k+{2\over z-2}\right)
     \over\Gamma\left(n+{2\over z-2}\right)}u_{n+k}(0)s^k
   \ea
Because of the asymptotic relation $x^{b-a} \Gamma(x+a)/\Gamma(x+b) = 1 +{\rm O}(1/x)$ \cite[Eq. (6.1.47)]{Abra}, in the
limit $z\to 2$ one recovers the result (\ref{Soluns}) or the linear chain. 
The final density of the $n$-string is obtained after performing the
transformations (\ref{RenorTime}) and (\ref{Trick1Bethe}) and since $C_0(0)=u_n(0)$, one arrives at eq.~(\ref{68}) in the main text.

\appsection{D}{Number of stationary states in the two-species model}

Generalising the discussion of the single-species model \cite{Carl01,deSm02,Henk04,Henk08}, 
we briefly outline the analogous computation
of the number of stationary states on a chain of $L$ 
sites in the two-species pair-annihilation model of section~4. 
The counting is based on the observation that stationary states cannot contain neither 
$AB$ nor $BA$ pairs on nearest-neighbour sites. 

First, consider an \underline{\em open chain}. Let $S_L$ 
be the number of stationary states for a chain of $L$ sites, and let
$A_L := \vec{N}\left(A\fbox{L-1}\right)$ denote the number of stationary states on a chain of 
$L$ sites where the leftmost site
is occupied by an $A$-particle. Similarly define $B_L = \vec{N}\left(B\fbox{L-1}\right)$ and 
$O_L := \vec{N}\left(\emptyset\fbox{L-1}\right)$, such that 
$S_L=A_L+B_L+O_L$. For small lattices, one has
$S_0:=1$ and $S_1=3$, $S_2=7$, $S_3=17$, $S_4=41$ \ldots. The three contributions to 
$S_L$ satisfy the recursions
\BEA
\hspace{-2.2truecm}O_L&\hspace{-1.5truecm}=&\vec{N}\left(\emptyset\fbox{L-1}\right) 
\:=\: \vec{N}\left(\emptyset\emptyset\fbox{L-2}\right) 
+ \vec{N}\left(\emptyset A\fbox{L-2}\right)+\vec{N}\left(\emptyset B\fbox{L-2}\right) 
\nonumber \\
&\hspace{-1.5truecm}=&O_{L-1} + A_{L-1} + B_{L-1} \:=\:  S_{L-1} 
\\
\hspace{-2.2truecm}A_L &\hspace{-1.5truecm}=& \vec{N}\left(A\fbox{L-1}\right) 
\:=\: \vec{N}\left(A\emptyset\fbox{L-2}\right) 
+ \vec{N} \left(AA\fbox{L-2}\right) 
= O_{L-1} + A_{L-1} 
\\
\hspace{-2.2truecm}B_L &\hspace{-1.5truecm}=& \vec{N}\left(B\fbox{L-1}\right) 
\:=\: \vec{N}\left(B\emptyset\fbox{L-2}\right) 
+ \vec{N} \left(BB\fbox{L-2}\right) 
= O_{L-1} + B_{L-1} 
\EEA
and collecting terms, we have the recursion
\BEQ
\hspace{-2.2truecm}S_L = 2 S_{L-1} + O_{L-1} = 2 S_{L-1} + S_{L-2}
\EEQ
The solution is readily found, for all $L\geq 0$ 
\BEQ
\hspace{-2.2truecm}S_L = \demi \left( 1 + \sqrt{2\,}\,\right)^{L+1} 
+ \demi \left( 1 - \sqrt{2\,}\,\right)^{L+1} 
\stackrel{L\gg 1}{\simeq} 1.207 \left( 1+\sqrt{2\,}\,\right)^L \,.
\label{A5}
\EEQ

Second, consider a \underline{\em periodic chain}. The number of stationary states is denoted by
$\overline{S}_L=\overline{A}_L +\overline{B}_L +\overline{O}_L$, where 
$\overline{A}_L := \vec{N}\left(\left.A\fbox{L-1}\right\|\right)$ 
denotes the number of stationary states on a periodic chain of
$L$ sites where one of the particles is of species $A$ 
and the trait indicates the periodic boundary conditions. 
$\overline{B}_L$ and $\overline{O}_L$ are defined similarly. For small lattices, one  has
$\overline{S}_1=3$, $\overline{S}_2=7$, $\overline{S}_3=15$, $\overline{S}_4=35, \ldots$.  
Now, if one of the sites is empty, this opens the chain so that
the number of stationary states is equal to the one of an open chain of 
$L-1$ sites, hence $\overline{O}_L=S_{L-1}=O_{L}$. Furthermore
\BEA
\overline{A}_L &=& \vec{N}\left(\left.A\fbox{L-1}\right\|\right) 
\:=\: \vec{N}\left(\left.A\fbox{L-2}\,\emptyset\right\|\right)+ \vec{N}\left(\left.A\fbox{L-2}A\right\|\right) 
\nonumber \\
&=& \vec{N}\left(\left.\emptyset A\fbox{L-2}\right\|\right)+ \vec{N}\left(\left.AA\fbox{L-2}\right\|\right) 
\nonumber \\
&=& A_{L-1} + \vec{N}\left(\left.A\fbox{L-2}\right\|\right) \:=\: A_{L-1} + \overline{A}_{L-1}
\EEA
since for the counting of the stationary states the pair 
$AA$ acts in the same way as a single $A$-particle. Similarly, 
$\overline{B}_L = B_{L-1} + \overline{B}_{L-1}$. Adding these contributions, we find
\BEA
\overline{S}_L &=& S_{L-1} + A_{L-1} + B_{L-1} + \overline{A}_{L-1} + \overline{B}_{L-1} \nonumber\\
&=& \overline{S}_{L-1} + 2 \left( S_{L-1} - S_{L-2} \right) \nonumber \\
&=& 1+ \left( 1+\sqrt{2\,}\,\right)^L + \left( 1-\sqrt{2\,}\,\right)^L 
\stackrel{L\gg 1}\simeq \left( 1+\sqrt{2\,}\,\right)^L 
\label{A7}
\EEA
where the expression in the last line, valid for $L\geq 1$, is obtained by using (\ref{A5}). 

As found previously for the single-species model \cite{Carl01,Dutta09,Henk08}, 
the number of stationary states is on an open chain about
$20\%$ larger than for a periodic chain. 



\newpage 
\begin{thebibliography}{999}

\bibitem{Abra} M. Abramovitz and I.A. Stegun, {\it Handbook of mathematical functions}, Dover (New York 1965)

\bibitem{Abad04} E. Abad, Phys. Rev. {\bf E70}, 031110 (2004).

\bibitem{Agha05} A. Aghamohammadi and M. Khorrami, Eur. Phys. J. {\bf B47}, 583 (2005). 




\bibitem{Bale75} R. Balescu, {\it Equilibrium and non-equilibrium statistical mechanics}, Wiley (New York 1975). 

\bibitem{Baxter82} R.J. Baxter, 
{\it Exactly solved models in statistical mechanics}, Academic Press (London 1982).

\bibitem{benA92} D. ben Avraham and J. K\"ohler, Phys. Rev. {\bf A45}, 8358 (1992). 

\bibitem{benA98} D. ben Avraham, Phys. Rev. Lett. {\bf 81}, 4756 (1998). 

\bibitem{benA00} D. ben Avraham and S. Havlin, 
{\it Diffusion and reactions in fractals and disordered systems},
Cambridge University Press, (Cambridge 2000)

\bibitem{benA07} D. ben Avraham and M.L. Glasser, J. Phys. Cond. Matt. {\bf 19}, 065107 (2007). 

\bibitem{Bonn95} B. Bonnier and E. Pommiers, Phys. Rev. {\bf E52}, 5873 (1995)

\bibitem{Bonn00} B. Bonnier, Phys. Rev. {\bf E58}, 5424 (1998); Phys. Rev. {\bf E61}, 1270 (2000). 

\bibitem{Brey03} J. Brey and A. Prados, Phys. Rev. {\bf E68}, 051302 (2003). 

\bibitem{Brey07} J. Brey and A. Prados, Powder Technology {\bf 182}, 272 (2007). 

\bibitem{Burl87} S.F. Burlatsky and A.A. Ovchinnikov, Sov. Phys. JETP {\bf 65}, 908 (1987).

\bibitem{Burl90} S.F. Burlatsky and A.I. Chernoustsan, Phys. Lett. {\bf A145}, 56 (1990). 

\bibitem{Carl01} E. Carlon, M. Henkel and U. Schollw\"ock, Phys. Rev. {\bf E63}, 036101 (2001). 
 
\bibitem{Coul04} C. Coulon, R. Cl\'erac, L. Lecren, W. Wernsdorfer and H. Miyasaka, 
Phys. Rev. {\bf B69}, 132408 (2004).

 

\bibitem{Dauch07} O. Dauchot, in M. Henkel, M. Pleimling and R. Sanctuary (eds), 
{\it Ageing and the glass transition},
ch. 4, p. 161,  Springer Lecture Notes in Physics {\bf 716}, Springer (Heidelberg 2007). 

\bibitem{deSm02} G. de Smedt, C. Godr\`eche and J.-M. Luck, Eur. Phys. J. {\bf B27}, 363 (2002). 

\bibitem{Deker79} U. Deker and F. Haake, Z. Phys. {\bf B35}, 281 (1979). 

\bibitem{Doer90} C.R. Doering and M. Burschka, Phys. Rev. Lett. {\bf 64}, 245 (1990). 

\bibitem{Dura10} X. Durang, J.-Y. Fortin, D. del Biondo, M. Henkel and J. Richert, 
J. Stat. Mech. P04002 (2010). 

\bibitem{Dura11} X. Durang, J.-Y. Fortin and M. Henkel, J. Stat. Mech. P02030 (2011).

\bibitem{Dutta09} S.B. Dutta, M. Henkel and H. Park, J. Stat. Mech. P02023 (2009). 


\bibitem{Edwa94} S.F. Edwards, in A. Mehta (ed) 
{\it Granular Matter: an interdisciplinary approach}, Springer (Heidelberg 1994). 

\bibitem{Evan84} J.W. Evans, J. Math. Phys. {\bf 25}, 2527 (1984). 

\bibitem{Evan93} J.W. Evans, Rev. Mod. Phys. {\bf 65}, 1281 (1993). 

\bibitem{Fred84} H.G. Frederikson H.C. Andersen Phys. Rev. Lett. {\bf 53}, 1244 (1984). 

\bibitem{Glauber63} R. Glauber, J. Math. Phys. {\bf 4}, 263 (1963). 

\bibitem{Godr05} C. Godr\`eche and J.-M. Luck, J. Phys.: Condens. Matter {\bf 17}, S2573 (2005). 

\bibitem{Haake80} F. Haake and K. Thol, Z. Phys. {\bf B40}, 219 (1980). 


\bibitem{Henk04} M. Henkel and H. Hinrichsen, J. Phys. {\bf A37}, R117 (2004). 



\bibitem{Henk08} M. Henkel, H. Hinrichsen and S. L\"ubeck, 
{\it Non-equilibrium phase transitions Vol. 1: absorbing phase transitions}, Springer (Heidelberg 2008)

\bibitem{Kamke59} E. Kamke, {\it Differentialgleichungen: L\"osungsmethoden und L\"osungen}, 
vol. 2, 4$^{\rm th}$ edition, Akademische Verlagsgesellschaft (Leipzig 1959).

\bibitem{Kapr92} P.L. Kaprivsky, Chem. Phys. {\bf 168}, 15 (1992). 

\bibitem{Kemk81} V.M. Kemkre and H.M. van Horn, Phys. Rev. {\bf A23}, 3200 (1981). 

\bibitem{Khor12} M. Khorrami and A. Aghamohammadi, Eur. Phys. J. {\bf B85}, 134 (2012).

\bibitem{Kimball79} J.C. Kimball, J. Stat. Phys. {\bf 21}, 289 (1979).

\bibitem{Kirk35} J.G. Kirkwood, J. Chem. Phys. {\bf 3}, 300 (1935). 

\bibitem{Kope90} R. Kopelman, C.S. Li and Z.-Y. Shi, J. Luminescence {\bf 45}, 40 (1990). 

\bibitem{Kroo93} R. Kroon, H. Fleurent and R. Sprik, Phys. Rev. {\bf E47} 2462 (1993). 

\bibitem{Kuzo88} V. Kuzovkov and E. Kotomin, Prog. Phys. {\bf 51}, 1479 (1988). 

\bibitem{Ludi91} S. Luding, H. Sch\"orer, V. Kuzovkov, A. Blumen, J. Stat. Phys. {\bf 65}, 1261 (1991). 

\bibitem{Maju93} S.N. Majumdar and V. Privman, J. Phys. {\bf A26}, L743 (1993).

\bibitem{Marr99} J. Marro and R. Dickman, {\it Nonequilibrium phase transitions in lattice models}, 
Cambridge University Press (Cambridge 1999). 


\bibitem{Oliv92} M.J. de Oliveira, T. Tom\'e and R. Dickman, Phys. Rev. {\bf A46}, 6294 (1992). 

\bibitem{Oliv94} M.J. de Oliveira and T. Tom\'e, Phys. Rev. {\bf E50}, 4523 (1994). 

\bibitem{Prad01} A. Prados and J. Brey, J. Phys. {\bf A34}, L453 (2001). 

\bibitem{Prad02} A. Prados and J. Brey, Phys. Rev. {\bf E66}, 041308 (2002). 

\bibitem{Pras89} J. Prasad and R. Kopelman, Chem. Phys. Lett. {\bf 157} (1989) 535

\bibitem{Priv97} V. Privman (ed), {\it Non-equilibrium statistical mechanics in one dimension}, 
Cambridge University Press (Cambridge 1997). 

\bibitem{Rabe11} M. Rabe, D. Verdes and S. Seeger, Adv. Coll. Interface Sci. {\bf 162}, 87 (2011). 


\bibitem{Russ06} R.M. Russo, E.J. Mele, C.L. Kane, I.V. Rubtsov, M.J. Therien and D.E. Luzzi, Phys. Rev.
{\bf B74} (2006) 041405(R)

\bibitem{Scha00} P. Schaaf, J.C. Voegel and B. Senger, J. Phys. Chem. {\bf B104}, 2204 (2000). 

\bibitem{Schn89} H. Sch\"orer, V. Kuzovkov, A. Blumen, Phys. Rev. Lett. {\bf 63}, 805 (1989)

\bibitem{Schn90} H. Sch\"orer, V. Kuzovkov, A. Blumen, J. Chem. Phys. {\bf 92}, 2310 (1989)

\bibitem{Schu00} G.M. Sch\"utz, in C. Domb and J.L. Lebowitz (eds), 
{\it Phase transitions and critical phenomena}, Vol. 19, p. 1, Academic Press (London 2000). 

\bibitem{Sriv09} A. Srivastava and J. Kuno, Phys. Rev. {\bf B79} (2009) 205407


\bibitem{Tarj04} G. Tarjus and P. Viot, Phys. Rev. {\bf E69}, 011307 (2004). 

\bibitem{Tome01} T. Tom\'e and M.J. de Oliveira, 
{\it Din\^amica estoc\'astica e irreversibilidade}, 
Editora da Universidade de S\~ao Paulo (S\~ao Paulo 2001). 

\bibitem{Tous83} D. Toussaint and F. Wilczek, J. Chem. Phys. {\bf 78} 2642 (1983). 

\end{thebibliography}
\end{document}